# Emergent scattering regimes in disordered metasurfaces near critical packing

M. Chen[1], A. Agreda[1‡], T. Wu[1], F. Carcenac[2], K. Vynck[3], P. Lalanne[1*]

[1]LP2N, CNRS, IOGS, Université Bordeaux, Talence, France

[2]LAAS-CNRS, Université de Toulouse, CNRS, Toulouse, France

[3]Institut Lumière Matière, CNRS, Université Claude Bernard Lyon 1, Villeurbanne, France

[‡]Current address: ELORPrintTec, Allée Geoffroy Saint-Hilaire, F-33600 Pessac, France

[*]Corresponding author: Philippe.Lalanne@institutoptique.fr

Abstract: Disordered metasurfaces provide a versatile platform for harnessing near- and far-field scattered light. Most research has focused on either particulate topologies composed of individual, well-identified metaatoms or, to a lesser extent, semi-continuous aggregate topologies without well identified inclusions. Here, we uncover an intermediate 'critical packing' regime characterized by metasurface morphologies in which a significant fraction of metaatoms begin to connect. We experimentally demonstrate that, at this threshold, the properties of the scattered light abruptly change and, via a statistical quasinormal mode analysis, interpret this change as a marked transition in the statistics of the photon density of states. Unlike percolation in semicontinuous metal films, this transition affects not only the specular but also the diffuse components of the scattered light in a profound way. Our results introduce critical packing topologies as a novel design strategy for manipulating the spectral and angular characteristics of light using ultrathin optical coatings. Emergent functionalities include colour shifts in diffuse light driven by multiple scattering and surface whitening, with potential applications in display technologies—for example, to reduce glare in electronic screens.

## Introduction

In solid-state systems, rapid changes in the electronic density of states (DoS) are often associated to spectacular phenomena, such as phase transitions [1-3]. In random photonic materials, the analogue of the electronic DoS is played by the photon density of states (PDoS) and the states are quasinormal modes (QNMs) with complex energies, owing to non-Hermiticity induced by leakage or absorption [4]. The PDoS contains crucial information on the optical transport properties of the material [5-11], with important implications on light localization [12-14], bandgap formation [15,16] or lasing in the presence of gain with long-lived quasi-extended modes [17]. In this work, we report a phase-transition-like phenomenon in the context of dielectric disordered metasurfaces, revealing a close photonic analogue to some behaviour observed in electronic systems.

Optical metasurfaces are artificial surfaces in which the constituent inclusions (or metaatoms) are meticulously engineered from dielectric or metallic subwavelength elements to achieve tailored optical properties and functionalities [4,19]. Among these, disordered metasurfaces hold a prominent place, both from fundamental [20,21] and applied [22,23] perspectives. They have led to new designs for optical encryption [24,25], light extraction [26,27], transparent displays [28], low-emissivity coatings [29], non-iridescent colouration [30-33], visual appearance design [34], and optical diffuser [35], among others.

This work introduces disordered metasurfaces operating at the transition between particulate and aggregate topologies, where metaatoms begin to connect (Fig. 1). In this unexplored regime, referred to as critical packing hereafter, we theoretically predict a sudden transition in the statistical distribution of QNMs when morphological parameters like the metaatom density or spatial correlation are varied. We experimentally verify that this transition results in significant changes in the behaviours of both specular and diffuse light, offering new opportunities for creating ultrathin fade-resistant coatings designed to implement advanced functionalities, unattainable through multilayer thin films [22,23,30-34,37-38].

The phase transition in our disordered metasurfaces shares similarities with the well-known system of semicontinuous metal films [39,40], yet it also reveals significant differences. Semicontinuous metal films consist of deeply subwavelength (typically $< 20$ nm) nanoislands, which exhibit much larger absorption cross-sections than scattering cross-sections and are arranged with limited short-range order. In contrast, our metasurfaces exhibit a different hierarchy: scattering dominates over absorption, and we achieve highly precise short-range spatial correlations. These differences help explain why semicontinuous metal films are predominantly studied for their absorption properties and specular responses [23], while in our metasurfaces, the spectral and angular behaviour of diffuse light plays a central role. Furthermore, the optical properties of semicontinuous metal films are often analysed through a geometric phase transition, where small, disconnected clusters merge into larger, connected clusters—ultimately reaching a percolation threshold that spans the entire system. In our metasurfaces, we also observe a geometric phase transition; however, here, the transition occurs when the metaatoms connect locally, forming small clusters well before a percolation threshold with spanning clusters is reached.

Hereafter, we focus on a set of disordered metasurfaces consisting of silicon resonant metaatoms on an optically inert glass substrate. This simple system is intentionally selected to isolate the effects that arise near the transition between particulate and aggregate topologies only. As illustrated in Figure 1, we exploit three key control knobs — metaatom size, density, and packing fraction — of particulate metasurfaces to gradually explore the transitioning between particulate and aggregate structures. The central idea of this work is to examine the sudden changes in the optical response of metasurfaces around critical packing by tuning these variables.

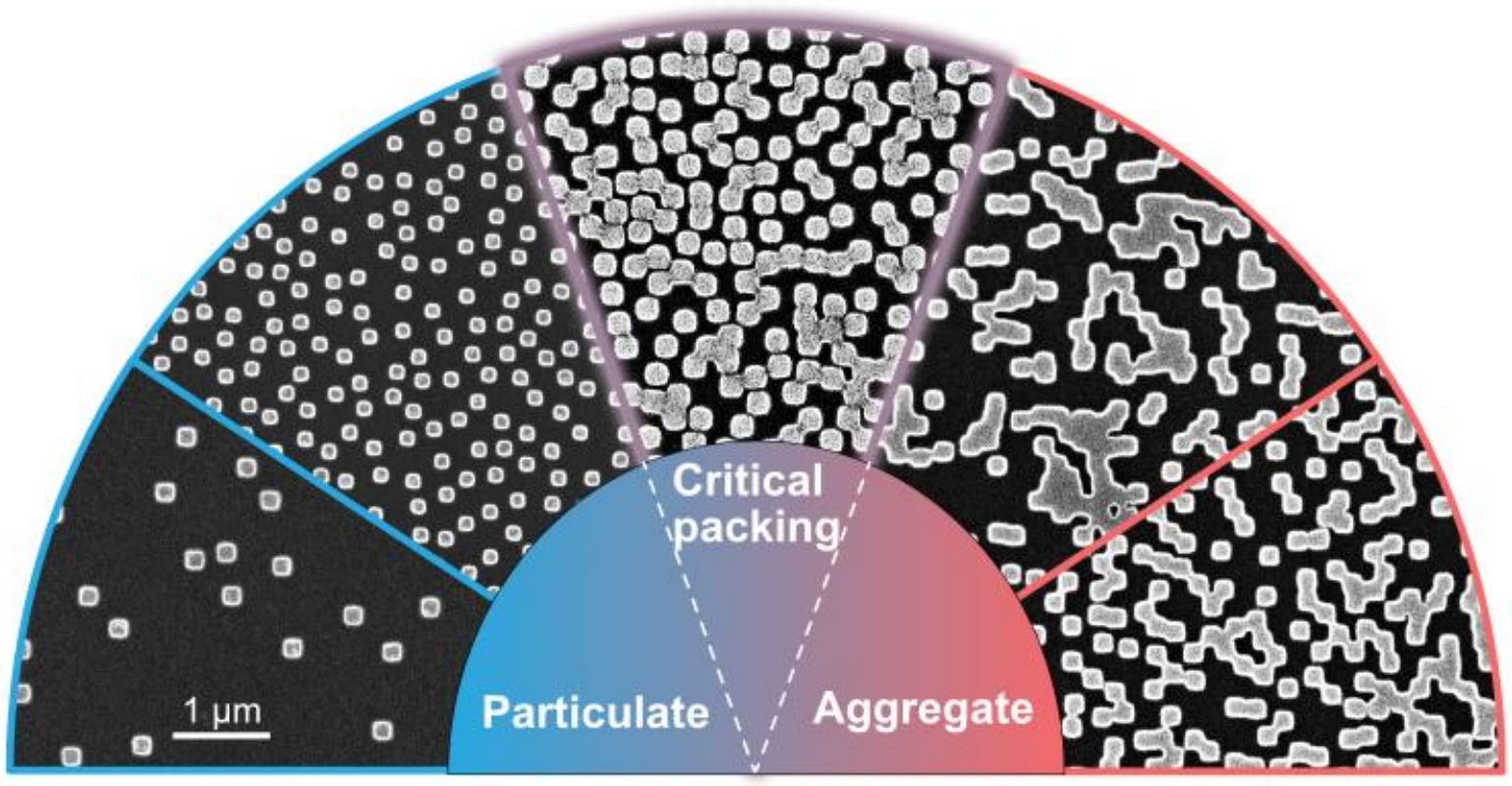

**Figure 1**. **Metasurfaces transitioning around critical packing.** The figure illustrates the transition from particulate (blue) to semicontinuous aggregate (red) topologies through critical packing topologies (purple), where approximately half of the nanoparticles touch each other. All metasurfaces are created using electron beam lithography by writing disordered arrays of squares by adjusting three key fabrication knobs: the size, density, and spatial correlation of the squares (see Method).

## Manifold visual appearances around critical packing

To experimentally investigate the light scattering transition between particulate and aggregate topologies, we designed a series of disordered metasurfaces written in a negative resist using electron beam lithography and then transferred into a 145 nm-thick silicon layer on a glass substrate (Fig. 2a and Methods). The metasurfaces are circular with a 300 µm diameter. They are fabricated on the same sample to reduce fabrication variability. All the electron beam patterns consist of random arrays of square features, whose centres are positioned using a Poisson disk sampling algorithm [41]. The $(x, y)$ coordinates of the centres are formatted into Graphic Data System (GDS) files.

We vary three control knobs of the square features, their density, size and disorder correlation. Structural correlation is controlled by setting a minimum interparticle distance, $a$, between particle centres, resulting in an effective surface coverage or 'packing fraction' defined as $p = \rho\pi a^2/4$.

This process yields a comprehensive set of 27 metasurfaces. The designed patterns feature squares with lateral sizes $\ell = 95$, 130 and 170 nm, densities $\rho = 2$, 5 and 10 $\mu\text{m}^{-2}$, and packing fractions $p =$ 0.1, 0.3, and 0.5. These values represent nominal target parameters, but different topological states emerge during fabrication due to spatial overlaps between the square features at certain densities, sizes, or disorder correlations. Scanning electron microscope (SEM) images of all metasurfaces are shown in Fig. 2c, with a few previously displayed in Fig. 1. The metasurfaces can be categorized into three topological states: particulate, aggregate, and critical-packing topologies, where particles begin to touch.

Both the specular and diffuse reflection responses of the metasurfaces are measured using a gonio-spectrometer setup and a slightly focused supercontinuum laser. After normalizing the measurements with the diffuse response of a reference diffuser, which offers high diffuse and Lambertian reflectance, we derive the bidirectional reflectance distribution function (BRDF) [42]. This multidimensional function characterizes how the metasurfaces scatter light across all viewing directions, depending on wavelength, incident angle and polarization. In addition to these quantitative measurements, the light scattered off the metasurfaces is visualized on a centimetre-scale hemispheric diffusive screen.

A halved ping pong ball serves as the screen, with an elongated band apertures along the plane of incidence to prevent blocking the incident beams and their specular reflections (Fig. 2b). This simple setup provides a direct visualization of the colour and scattered intensity produced by the metasurfaces, offering a qualitative representation of their visual appearances in reflection. Photographs of the ping pong ball, capturing the diffuse light from all 27 metasurfaces, are taken using a Canon EOS 1000D DSLR. Specular reflection is also photographed by aligning the camera with the specular direction. Different camera settings are used for specular and diffuse light images to prevent saturation, but the settings remain consistent across all recordings, allowing for direct intercomparison of the images.

The 27×2 photographs capturing both the diffuse components of scattered light at 45° incident angle and the specular components at a 10° incident angle under unpolarized light illumination are displayed in Fig. 2c. They are presented alongside corresponding SEM images of the metasurfaces, with the specular light photographs appearing as rectangular colour boxes positioned just to the left of the SEM images.

Particulate metasurfaces are primarily located in the leftmost column and the upper portions of the central and rightmost columns of Fig. 2c. With the exception of the singular metasurface in the upper right corner, which will be discussed separately, they exhibit two dominant, vivid hues, primarily determined by the resonance frequency of individual metaatoms. For uncorrelated disorder, $p = 0.1$, the colour and brightness of the diffuse light remain independent of the observation direction. In contrast, for correlated disorder, $p = 0.5$ , brightness is reduced around the specular direction.

This trend is a direct manifestation of the engineered correlations, quantitatively captured by the static structure factor, $S_r$ (see Suppl. Section S1 and Fig. S1d and [23]). For weakly correlated patterns ($p = 0.1$), $S_r$ is nearly constant, resulting in the observed uniform, isotropic scattering. As correlation increases (e.g., $p = 0.5$), a pronounced dip develops in $S_r$ at small in-plane wavevectors ($q_{||} \approx 0$), which directly causes the reduction of diffuse light around the specular direction. Concurrently, a strong peak emerges for $q_{||} \approx 2\pi\sqrt{\rho}$, creating the bright, ring-shaped halo of enhanced scattering. The dispersive nature of this peak, where its angular position depends on wavelength, is what gives the halo its distinct colour. Furthermore, as the particle density $\rho$ increases from 2 to 5 $\mu\text{m}^{-2}$ (moving from the left to the middle column in Fig. 2c), the primary peak of the structure factor shifts to shorter wavelengths (Suppl. Fig. S1b). This shift directly explains why the coloured halo appears to expand and change hue, providing another knob for tuning the visual appearance [34]. These observations are further analysed in Suppl. Section S1, which presents additional variations in vivid hues obtained by adjusting the metaatom size parameter.

Metasurfaces operating near critical packing appear in the middle and lower sections of the second and third columns. They display pastel hues with low saturation, incorporating a significant amount of white light. Some of these metasurfaces, particularly at packing fractions of 0.5 and 0.3, exhibit an angle-dependent brightness, with a noticeable reduction of diffuse light around the specular direction. This confirms that correlation in the initial electron-beam pattern continues to affect scattering properties, even as the metaatom overlap is significant.

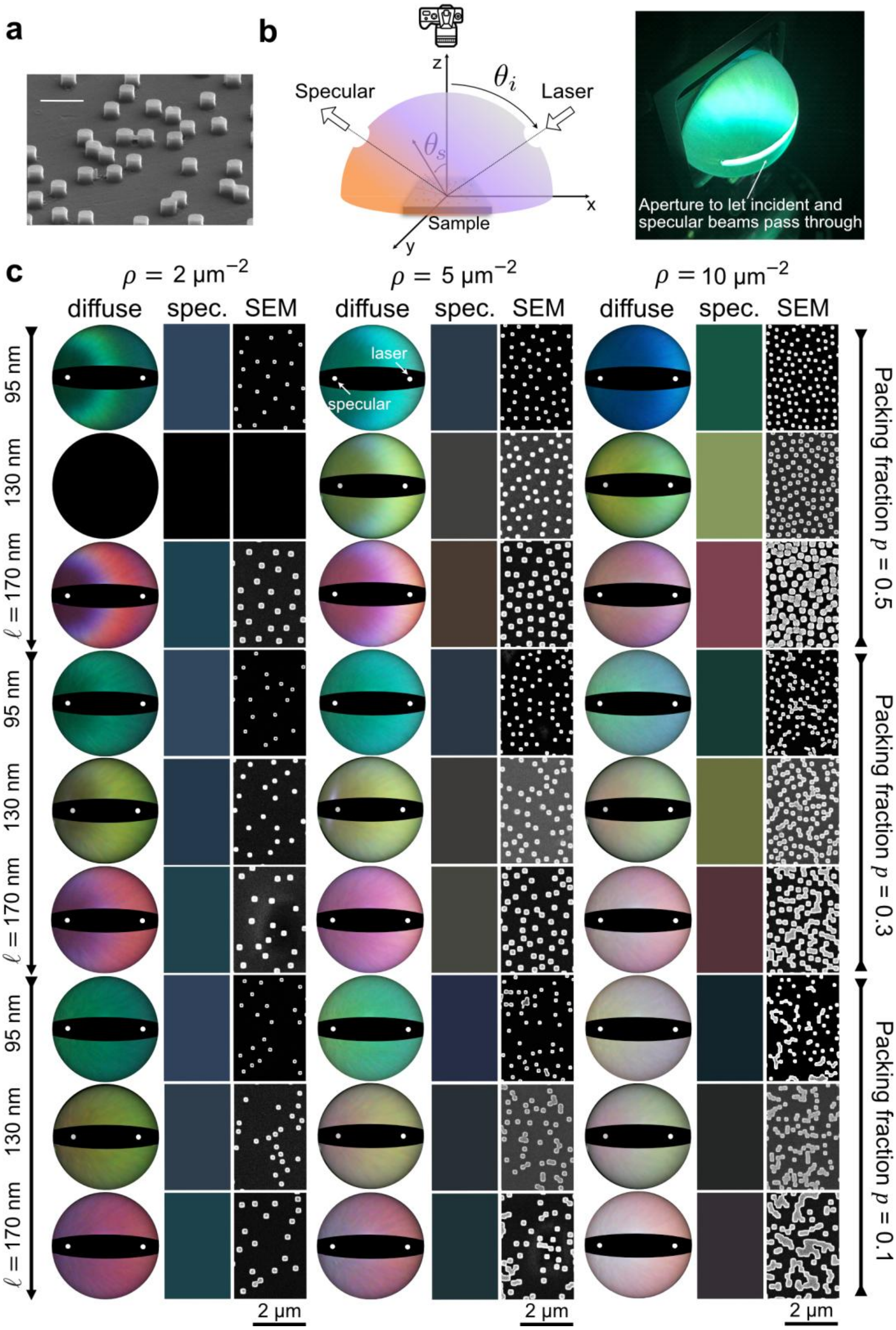

**Figure 2**. A great variability of visual diffuse and specular responses is obtained with a restricted set of manufacturing parameters. **a** SEM image of a metasurface. The scale bar represents 500 nm. **b** Ping-pong ball setup used to visualize the diffuse component of the BRDF upon illumination with a focused supercontinuum laser beam. **c** Photographs of the ping-pong ball show the diffuse responses for 27 metasurfaces under 45° incident illumination, with nominal square features of side length $\ell = 95$, 130 and 170 nm, densities $\rho = 2$, 5 and 10 $\mu\text{m}^{-2}$, and packing fractions $p = 0.1$, 0.3 and 0.5. The two white dots indicate the incident and specular directions. Specular colours recorded at 10° incidence for the metasurfaces are displayed as uniformly coloured rectangles. SEM images depict the arrangement, size, and density of the metaatoms. Note that the metasurface with $\ell = 130$ nm, $p = 0.5$ and $\rho = 2\ \mu\mathrm{m}^{-2}$ has been damaged during sample handling.

Aggregate metasurfaces, which form at high density and low packing fraction, are primarily found in the lower part of the third column. They are characterized by high lightness, close to white, with soft hues that slightly vary between the specular and backward directions. Unlike particulate metasurfaces, this angular-dependent colouring does not stem from correlated disorder, as aggregate metasurfaces emerge from a weakly correlated configuration and do not exhibit a reduction in diffuse light intensity around the specular direction. A closer inspection reveals pronounced shifts in hue as the system approaches and crosses critical packing, including both a progressive blue-shift and a general whitening of the diffuse light. These spectral trends are not immediately obvious in Fig. 2 alone, but will be quantitatively analysed in detail in Figs. 4 and 5.

Specular colours tend to be relatively dull for aggregate metasurfaces, as well as for particulate metasurfaces with low densities, $\rho = 2$ and $5\ \mu\mathrm{m}^{-2}$, where blue-green or brown hues dominate. However, bright colours emerge near critical packing at a density of 10 $\mu\mathrm{m}^{-2}$, and these colours differ significantly from those observed in diffuse light. This distinction arises from the fundamentally different physical processes governing the specular and diffuse components of the scattered light. Specular reflection is formally associated to the average field in the material, which propagates in an effective homogeneous medium and is therefore subject to classical thin-film interference effects. This can lead to significant variations of the specular colour with small changes of metasurface parameters and illumination conditions. Diffuse reflection, on the other hand, is associated to the fluctuations of the scattered field around the average. Diffuse colours are mostly driven by the individual scattering elements dressed by the interaction with their neighbours, namely, for particulate metasurfaces, by dressed Mie resonances and particle pair correlations. Colour variations with the incident or scattered angles are often milder for the diffuse component than for the specular one.

Overall, the photographs in Figs. 2c and S1 illustrate a striking diversity of visual appearances, including angular-dependent intensity, variations in lightness, whitening effects, distinct diffuse and specular colours, and changes in chroma as density and correlation vary. What makes this diversity particularly remarkable is the transition from vivid hues to softer tints near critical packing. Equally noteworthy is that this broad range of optical effects is achieved using simple, manufacturable samples on a neutral silica substrate. The fact that all samples share identical raw materials, uniform square-shaped e-beam patterns, and a fixed etching depth underscores the profound role of the nanoscale topology itself in generating such diverse appearances.

## Statistical quasinormal mode analysis around critical packing

Let us now examine the cause of the sudden changes in the specular and diffuse reflectance arising as metasurfaces transition around critical packing. By analogy with solid-state systems, we consider the PDoS. Because of leakage in the air clads and absorption in silicon, the QNM frequencies are complex valued and we thus define the PDoS as the normalized number of photon states available at a particular energy and given quality factor.

Modal analyses of waves in disordered systems often rely on point scatterers with electric dipole resonances to linearize the eigenvalue problem [43]. This restriction is completely relaxed hereafter by using full-field electromagnetic simulations and a supercell approach with Bloch boundary conditions in the lateral $x$ and $y$-directions and perfectly matched layers (PMLs) in the $z$-direction (see Methods). To ensure the accuracy of our predictions, we systematically test them by progressively increasing the supercell size.

Our smaller supercells consist of approximately 30 silicon metaatoms and have a lateral size of 1.2×1.2 $\mu\mathrm{m}^2$. The metaatoms side length is $\ell = 100$ nm and their height is 145 nm. All metasurface patterns are generated using a Poisson disc sampling algorithm [41] for a fixed filling fraction of 0.2. The QNMs are computed and normalized using the freeware MAN (Modal Analysis of Nanoresonators) [44]. For computational efficiency, we assume a fixed silicon permittivity of 20-0.5$i$, an average value derived from literature data [45] for the visible spectrum ($\lambda = 400$ to $750$ nm); the negative imaginary part signifies optical loss under the $e^{i\omega t}$ time-harmonic convention used by our solver. Additionally, we approximate the electromagnetically inert silica substrate as air, introducing a planar symmetry at $z = 0$, which reduces the computational time. Based on this assumption, we classify modes as even or odd according to the symmetry or asymmetry of the $x$-component of the QNM electric field. The PDoS maps are computed from a total of ≈ 7600 QNMs over 40 independent realizations, providing a statistically stable dataset. Additional QNM maps can be found in Supplementary Section S2.2.

To unify the varied structural configurations that arise from tuning the metaatom size, density, and correlation, we introduce an effective parameter, the metaatom overlap rate $\tilde{O}$ defined as the fraction of metaatoms which are touching at least one neighbour. This fraction is calculated directly from the random pattern generated for the electron beam writer with the Poisson disc sampling algorithm. The overlap rate provides a direct measure of the structural connectivity and serves as a fundamental variable to track the system's progression from the particulate ($\tilde{O} = 0$) to the aggregate regime.

Figure 3a summarizes our key computational results for the PDoS maps for various topologies. To provide a comprehensive picture, each map uses two distinct colour scales: the main heatmap visualizes the PDoS itself (right colour bar, logarithmic scale), while the horizontal colour bands above the maps represent the averaged QNM excitation probability (left colour bar, linear scale). The four panels show the progression through the different regimes: (i) particulate $\tilde{O} = 0\%$, (ii) and (iii) critical packing $\tilde{O} = 10\%$ and $27\%$, and (iv) aggregate $\tilde{O} = 49\%$. The corresponding packing fractions are (i) $p = 0.5$, (ii) $p = 0.2$, (iii) $p = 0.16$ and (iv) $p = 0.1$.

The spectral range covered by the maps corresponds to the visible region, within which the silicon subwavelength metaatoms exhibit several distinct low-loss resonances. These include a quadrupolar-electric (EQ) mode, a degenerate dipolar-electric (EDx and EDy) mode, and a dipolar-magnetic (MDz) mode. These modes are highlighted with bright cyan squares on the maps; the $x$-component of their electric fields in the $xy$-plane are shown in the black inset in Fig. 3b.

The maps in Fig. 3a reveal a clear redistribution of the QNMs in the complex frequency plane as the metasurfaces transition through critical coupling, from isolated clouds in (i) to a single, regular cloud in (iv).

**PDoS diversity.** For particulate metasurfaces (i), the PDoS map appears as a patchwork of distinct QNM clouds, each centred predominantly around the individual QNM frequencies. One prominent example is the high-$Q$ ($\sim 80$) cloud at $\ell/\lambda \sim 0.16$, which primarily results from the hybridization of MDz modes that mostly radiate in-plane [46]. A representative QNM from this group is Mode 1 in Fig. 3b(i). This mode exhibits a highly ordered field distribution localized in air, which is reminiscent of leaky guided resonances at $k_{\parallel} = 0$ in grating waveguides [47]. Additional QNM clouds emerge between the real frequencies of the individual modes due to hybridization between different mode types —for instance, the cloud spanning $0.17 < \ell/\lambda < 0.2$ is driven by the interaction between EDx and MDz modes.

A distinguishing feature of the particulate PDoS map is the emergence of pronounced pseudogaps, with a notable example at $\ell/\lambda \approx 0.21$. These gaps arise from destructive interference among waves scattered by the strongly correlated metaatoms, analogous to the formation of photonic bandgaps in crystals. The analogy is reinforced by the mode profiles: QNMs just below the gap localize their fields primarily within silicon (as in dielectric bands), whereas those just above the gap concentrate their fields in surrounding air (as in air band) [46].

At the other end, for aggregate topologies (iv), metaatoms organize into extended chains—comparable in size to the incident wavelength—that merge into irregular clusters. Due to substantial polydispersity in cluster size and morphology, the resulting QNMs are more sparsely and almost uniformly distributed across the visible spectrum. This leads to a single, broad QNM cloud with smaller $Q$-factors, in agreement with the fact that larger metaatoms are better radiators. This broad cloud contrasts sharply with the fragmented PDoS of particulate topologies and arises from the complete closure of the pseudogaps—a hallmark of the profound reorganization of the electromagnetic mode structure at critical packing. Typical QNM field profiles are shown in Fig. 3b(iv) as Mode 4 and Mode 5. Their electric-field distributions differ markedly from those of isolated Si nanopost resonances: instead of being confined within the posts, the fields form 'hot spots' predominantly along the boundaries of the irregular Si clusters.

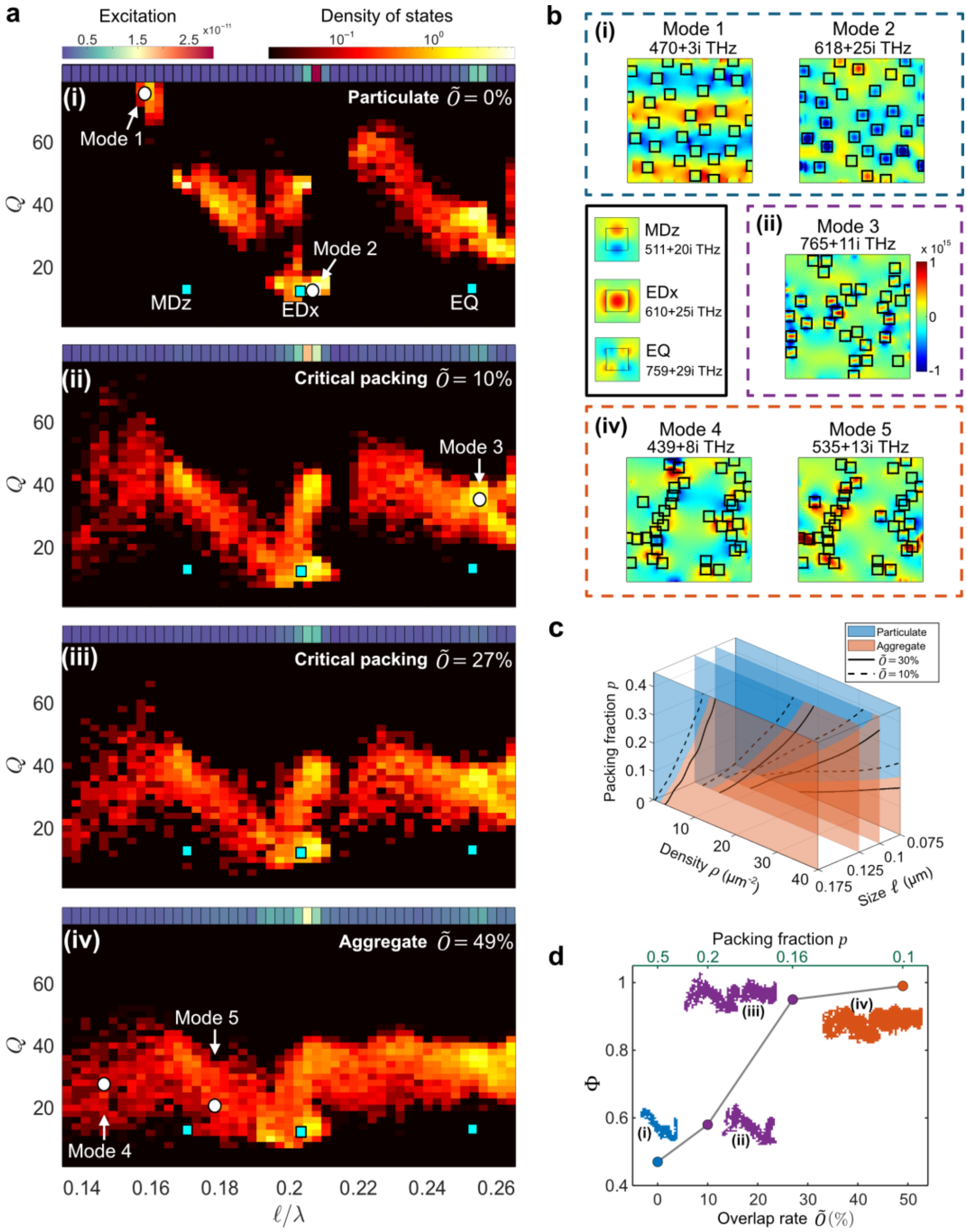

**Figure 3**. Evolution of the photon density of states (PDoS) when transitioning through critical-packing topologies. All metasurface patterns are generated using a Poisson disc sampling algorithm [41] for Si metaatoms with lateral size $\ell = 100$ nm and height of 145 nm. The filling fraction is 0.2. **a** PDoS maps of even electromagnetic QNMs for increasing overlap rate $\tilde{O}$: (i) particulate ($\tilde{O} = 0\%$), (ii–iii) critical packing ($\tilde{O} = 10\%$ and $27\%$), and (iv) aggregate ($\tilde{O} = 49\%$). The corresponding packing fractions are $p = 0.5$, $0.2$, $0.16$ and $0.1$, respectively. Colour bands above each map represent the averaged QNM excitation coefficients (see Methods), while cyan squares mark the dominant single-metaatom modes—MDz, EDx, and EQ. **b** Representative field maps showing the $x$-component of the QNM electric field (see Fig. S2.8 for details). The black inset shows the MDz, EDx and EQ QNM fields in the $xy$-plane. **c** $\rho - p - \ell$ phase diagram distinguishing particulate (blue) and aggregate (red) topologies. **d** Quantitative analysis of the PDoS evolution shown in **a**, from fragmented clouds at small $\tilde{O}$ values to a unique spanning cloud at $\Phi \approx 1$. Insets show the largest connected component for every PDoS map.

At critical packing, (ii) and (iii), the PDoS exhibits characteristics intermediate between the particulate and aggregate regimes. Even within the particulate topology, hybridized modes can shift by several linewidths from the resonance frequencies of isolated Si metaatoms, signalling the onset of strong inter-metaatom coupling. As the metaatoms begin to overlap, this coupling becomes the dominant factor shaping the PDoS, as seen in panels (ii) and (iii). In (ii), additional QNMs appear at lower frequencies, reflecting the formation of larger effective scatterers. In (iii), new hybridized modes emerge to fill the pseudogaps, marking the breakdown of the destructive interference that defines the particulate state.

Collectively, these features identify critical packing as the transitional regime where the PDoS undergoes a fundamental reorganization. The nature of this transitional regime is exemplified by Mode 3 in Fig. 3b(ii), which exhibits a hybrid character: on isolated metaatoms (left side of the unit cell), the internal fields show a high degree of phase coherence, with each particle exhibiting a similar field profile. In contrast, when the metaatoms begin to connect, this local order breaks down, and the field profile becomes non-uniform, signalling the onset of strong, disorder-induced hybridization.

**QNM excitation probability.** While the appearance of the PDoS map is informative, it is equally important to identify which modes are preferentially excited. The QNM excitation probability densities are represented by the colour bands above the PDoS maps in Fig. 3a and as colour dots in Figs. S2.2-S2.4, ranging from blue (weak excitation) to red (strong excitation). These probability densities are computed by averaging the QNM excitation coefficients over narrow spectral intervals (see Methods for the detailed procedure), thereby highlighting the spectral regions of the PDoS that are most effectively excited.

A striking feature of the particulate topology (i) is the strongly bimodal nature of the excitation distribution: only QNMs within two narrow frequency bands, near the EDx and EQ resonances, have significant excitation. At critical packing (ii) and (iii), this bimodality persists, though the peaks broaden and the maximum excitation strength diminishes. In the aggregate regime (iv), the excitation distribution becomes more heterogeneous, forming a multicoloured pattern that reflects generally weak-to-intermediate excitations across a broader frequency range. This trend aligns with the equipartition theorem [48,49], which suggests that, in highly disordered or chaotic systems, energy becomes uniformly distributed among all accessible degrees of freedom.

The near-field distributions in Fig. 3b provide further insight for these excitation trends. High cooperativity, where many metaatoms exhibit induced polarizations with similar phases, leads to the strong, coherent scattering required for efficient excitation by the incident plane wave. This is clearly seen in the highly-excited Modes 2. Conversely, the loss of long-range coherence in the aggregate regime, exemplified by the disordered field patterns of Modes 4 and 5, results in weaker, less efficient excitation. The hybrid nature of the critical packing modes, such as Mode 3, naturally corresponds to the intermediate excitation strengths observed during the transition.

Supplementary Section S2.3 examines the odd-symmetric modes, where $\tilde{E}_x(z) = -\tilde{E}_x(-z)$, and similarly identifies the critical packing transition as a pivotal point in the evolution of mode structure and excitation behaviour.

**Phase-transition diagram.** The appearance of the metasurfaces in Fig. 2 and the PDoS in Fig. 3a both suggest that the transition occurs when the metaatoms start to connect, irrespective of the specific trajectory chosen to transition, whether by varying $\rho, p$, or $\ell$. Consequently, the metaatom overlap rate $\tilde{O}$ is expected to be a universal, trajectory-independent parameter for tracking the phase change.

Figure 3c confirms this hypothesis. It provides a phase diagram that delineates the boundaries between particulate (blue) and aggregate (red) topologies across the full $\rho - p - \ell$ morphological parameter space. The critical-packing regime is highlighted by two constant-$\tilde{O}$ contours, corresponding to 10% (dashed) and 30% (solid) overlap rates. This phase diagram underscores that the critical-packing regime is trajectory independent and emerges as a general feature governed by the combined effects of all the morphological parameters.

To further reveal the impact of the transition in the PDoS maps, we further introduce the order parameter, $\Phi$, defined as the largest connected–component fraction. This metric, widely used in the statistical analysis of complex networks [50], measures the degree of fragmentation: $\Phi$ ranges from 0 to 1, with small values indicating highly fragmented networks.

Applied to probability density functions such as the PDoS, $\Phi$ is obtained by binarizing the PDoS maps and computing the ratio of the largest cloud area to the total area occupied by all clouds. In other words, $\Phi$ represents the fraction of "active" PDoS pixels that belong to the largest connected cloud [50]. Figure 3d plots $\Phi$ as a function of the overlap rate $\tilde{O}$, showing a sigmoidal evolution toward unity, with a sharp transition near $\tilde{O} = 20\%$. This evolution marks a shift from fragmented QNM clouds separated by bandgaps to a single, spanning cloud that extends across the entire visible spectrum. The coloured insets along the curve highlight this evolution, illustrating the largest connected region in each map.

Intuitively, as adjacent metaatoms begin to connect, the likelihood increases that additional metaatoms will also link up, leading to the rapid formation of larger clusters. This, in turn, results in a significant shift in PDoS connectivity. Thus, although Figs. 3a and 3d follow one specific trajectory in parameter space (with only $p$ varied), the abrupt transition in PDoS connectivity—and the associated pseudogap closure or optical response—should be considered a universal feature. This transition is expected to occur for any path that passes through the critical-packing regime. This universality is likely to hold when the optical wavelength is comparable to the metaatom size ($\lambda \sim n\ell$, where $n$ is the refractive index of the metaatoms), although the precise visual appearance (e.g., hue or saturation) may still depend on the specific trajectory taken.

## Blue shift of diffuse light in dense particulate metasurfaces

Strong modal coupling between localized resonances is a key factor driving the widespread use of subwavelength metaatoms [47], particularly when their optical near fields strongly overlap [51,52]. This principle is well established and widely exploited in ordered systems—such as small clusters of nanoresonators or periodic arrays—to finely tune the structural colours arising from collective resonances [53,54].

In contrast, the behaviour of disordered ensembles of nanoresonators remains less well understood. Due to the random spatial arrangement, the polarizations induced by multiple scattering fluctuate and tend to statistically average out [23,55], weakening coherent interactions. Consequently, strong coupling effects in disordered metasurfaces often manifest more subtly [55,56].

Figure 4 highlights a striking and unexpectedly blue shift that stands out as particularly remarkable in comparison. The shift is displayed for an incidence angle of 30°, with three ping-pong ball images obtained for $p = 0.5$ and increasing densities. At a low density of $\rho = 2\ \mu\mathrm{m}^{-2}$, multiple scattering is minimal. The diffuse light appears cadmium green, with an angle dependence hue marked by a suppressed diffusion near the specular reflection direction and a bright halo marked with a dashed arc line. These features—specular suppression and halo enhancement—are not central to the present discussion and are addressed in Supplementary Section S1.

As the density increases to 5 $\mu\mathrm{m}^{-2}$, the halo shifts to larger angles. The shift is accompanied by a modest increase in brightness and the onset of a perceptible bluing effect. At a high density of 10 $\mu\mathrm{m}^{-2}$ just below the critical packing threshold, a vivid sapphire blue emerges. Further increases in density would lead to a whitening of the diffuse light due to metaatom overlapping, as discussed in the next Section.

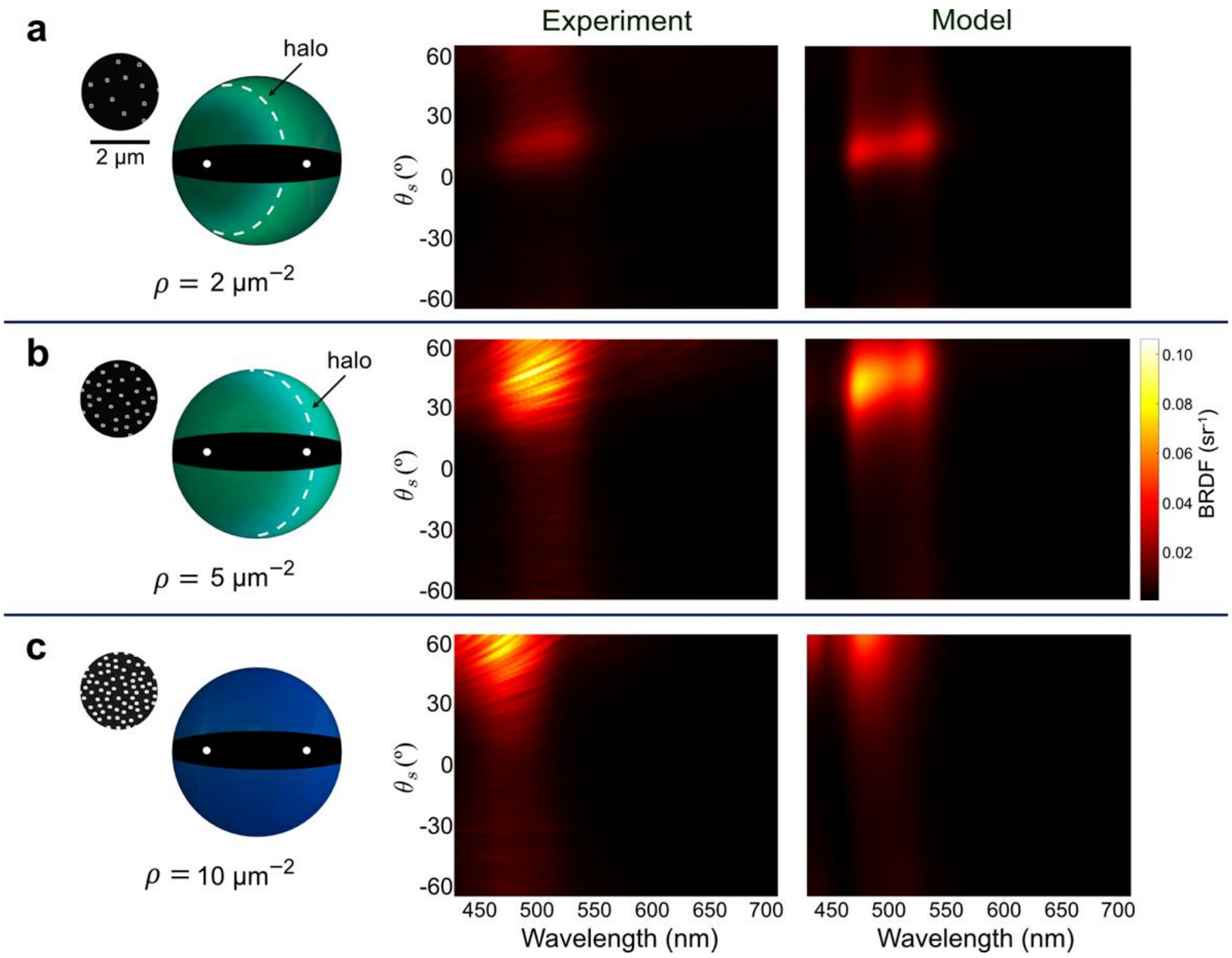

**Figure 4. Blue shift induced by multiple scattering just below critical packing threshold.** Three particulate metasurfaces are presented, each designed with the same metaatom size ($\ell = 95$ nm) and packing fraction ($p = 0.5$), but with increasing densities: **a** $\rho = 2\ \mu\text{m}^{-2}$; **b** $\rho = 5\ \mu\text{m}^{-2}$; **c** $\rho = 10\ \mu\text{m}^{-2}$. Measured and simulated in-plane BRDF maps are presented in the middle and right columns. The inclined streaks visible in the experimental data resemble speckle patterns, which arise from the small dimensions of the metasurfaces. Details of the algorithm employed to compute the correction factor can be found in Fig. S3. Note that $C(\mathbf{k}_\text{s}, \mathbf{e}_\text{s}, \mathbf{k}_\text{i}, \mathbf{e}_\text{i}) \approx 1$ for $\rho =$ 2 and 5 $\mu\text{m}^{-2}$. The correction factor matters only for the larger density, resulting in a blue shift of the BRDF peak. The ping-pong balls and the BRDF maps are collected at 30° incidence angle. Similar results are observed at 15° and 45° incidence angles.

Given that the shape and size of the metaatoms are precisely controlled during fabrication, the blue shift is attributed to direct electromagnetic interaction between metaatoms, particularly through evanescent and radiative couplings that are enhanced at higher densities. In most studies of cooperative scattering in particulate systems, particles are treated as point scatterers with resonant electric polarizabilities, where multiple scattering is modelled through dipole-dipole interactions, as seen in research on cold atomic clouds [57,58] and micro- and nanoscale discrete disordered media [59,60].

Instead, our high-index, 145nm-tall metaatoms support two electric dipolar modes, two magnetic dipolar modes, and one electric quadrupolar mode within the visible spectral range, see Fig. 3b. While it is possible to incorporate these multipoles using a closed set of equations [61], deriving a closed-form expression for diffuse light intensity that accounts for both multiple scattering and multipolar interactions is notoriously challenging, as it requires solving for the field-field correlation function [21].

To model the blue-shift effect, we employ a recent model that incorporates multiple scattering effects in a mean-field manner [34,37], without requiring explicit consideration of the multipolar response. In this model, the diffuse contribution to the BRDF is expressed as:

$$f_\text{diff} = \rho \frac{d\sigma_\text{s}}{d\Omega}(\mathbf{k}_\text{s}, \mathbf{e}_\text{s}, \mathbf{k}_\text{i}, \mathbf{e}_\text{i})\, S_r(\mathbf{k}_\text{s} - \mathbf{k}_\text{i})\, C(\mathbf{k}_\text{s}, \mathbf{e}_\text{s}, \mathbf{k}_\text{i}, \mathbf{e}_\text{i}), \quad (1)$$

where the first three terms — the density, the form factor $\frac{d\sigma_\text{s}}{d\Omega}$ which describes how a single isolated metaatom scatters light and the structure factor $S_r$ — neglect the electromagnetic interaction between metaatoms. Only the final term, the correction factor $C$, does account for this interaction and is therefore responsible for the observed blue shift. In Eq. (1), $\mathbf{k}_\mathrm{i}$ and $\mathbf{k}_\mathrm{s}$ denote the incident and scattered wavevectors and $\mathbf{e}_\mathrm{i}$ and $\mathbf{e}_\mathrm{s}$ represent the corresponding polarizations.

Without delving into details provided in Suppl. Section S3, the correction factor is theoretically inferred in the model from a mean-field estimation of the field driving every scatterer in the presence of multiple scattering. Through energy conservation arguments, it is possible to show that the sum of the specular transmittance ($T_\mathrm{s}$) and reflectance ($R_\mathrm{s}$) plays a key role to infer the mean field. In [34], these specular coefficients were inferred theoretically using advanced homogenization theories [62].

Here, we instead use measured values of $T_\mathrm{s}$ and $R_\mathrm{s}$, improving resilience to topological and material variations and minimizing potential inaccuracies in the homogenization theories. The spectral and angular maps of $(1 - T_\mathrm{s} - R_\mathrm{s})$, used to infer the correction factor, are shown in the insets of the rightmost column of Fig. 4. They are recorded with a 2-degree step variation in the incident angle and averaged over TE and TM incident light. Note that the maximum of $(1 - T_\mathrm{s} - R_\mathrm{s})$ is blue shifted for the denser metasurface for $\rho = 10\ \mu\mathrm{m}^{-2}$, compared to the lower densities. Consequently, the correction factor peaks near this wavelength (Suppl. Section S3), suggesting that the averaged local field exciting the metaatoms is maximized due to electromagnetic interaction.

To quantitatively assess the model accuracy, we measure the BRDF of the metasurfaces and compare the experimental data with theoretical predictions from Eq. (1). The intermediate steps for calculating the correction factor and subsequently deducing the BRDF are detailed in Fig. S3, which additionally includes colourmaps of $C(\mathbf{k}_\mathrm{s}, \mathbf{e}_\mathrm{s}, \mathbf{k}_\mathrm{i}, \mathbf{e}_\mathrm{i})$ as functions of wavelength and scattering angle.

The agreement between theory and experiment is notably good. The model effectively captures the BRDF increase and the resonant blue shift at high density. Remaining discrepancies likely stem from inaccuracies in modelling the metaatom shape and size, as well as moderate variations in these parameters, which significantly influence metaatom resonances.

The pronounced blue shift observed here is notably different compared to previously reported coupling effects induced by multiple scattering in high-density disordered metasurfaces [55,56]. This shift points to a novel mechanism for harnessing structural colouration in disordered metasurfaces, paving the way for innovative design strategies that move beyond conventional periodic architectures.

## Whitening effect across the critical packing transition

The previous effect arises for metasurfaces with tightly packed non-overlapping metaatoms. As the fabrication parameters are further tuned, critical packing threshold is reached, and the diffuse light shows important hue variations.

These variations are illustrated in Fig. 5a for a fixed metaatom size $\ell = 95$ nm. Two transitions from particulate to aggregate topologies are highlighted, either by decreasing the packing fraction from $p = 0.5$, to 0.3 and then 0.1 (transition (i)) or by increasing the density from $\rho = 2$, 5 and then 10 $\mu\mathrm{m}^{-2}$ (transition (ii)), causing the non-iridescent blue and green diffuse colours of the ping-pong ball images to converge into a white hue. The whitening is confirmed by in-plane BRDF measurements showing the spectral and angular dependence of the diffuse light for an incident angle of 30° (Fig. 5a). As we cross critical packing, the initial spectral distributions of the particulate topologies initially centred in the blue-green spectral region suddenly broaden, covering almost uniformly the entire spectral range from 420 nm to 700 nm.

To quantitatively access the purity of the white hue of the spectrally broad BRDF, we map the ping-pong ball images onto the CIE 1931 chromaticity diagram. This diagram is widely used to correlate reflection or transmission spectra with the colours perceived by human vision. The two transitioning are indicated by the trajectories (i) and (ii) in Fig. 5b, where we additionally mark the white point with coordinates (0.3127, 0.3290) with a blue circle. This point represents a standard reference for many colour applications with the D65 illuminant. Since our supercontinuum illuminant is different from the average daylighting, we also mark the white point with the D50 illuminant used in graphic arts, making it easier to evaluate the purity of the observed white, which is very close to the two marks. This trend is further quantified in Supplementary Section S6, where the 3D CIELAB representation confirms the trajectories' convergence toward the white point in chromaticity, while also explicitly showing the marked increase in lightness (L*) that accompanies the visual effect.

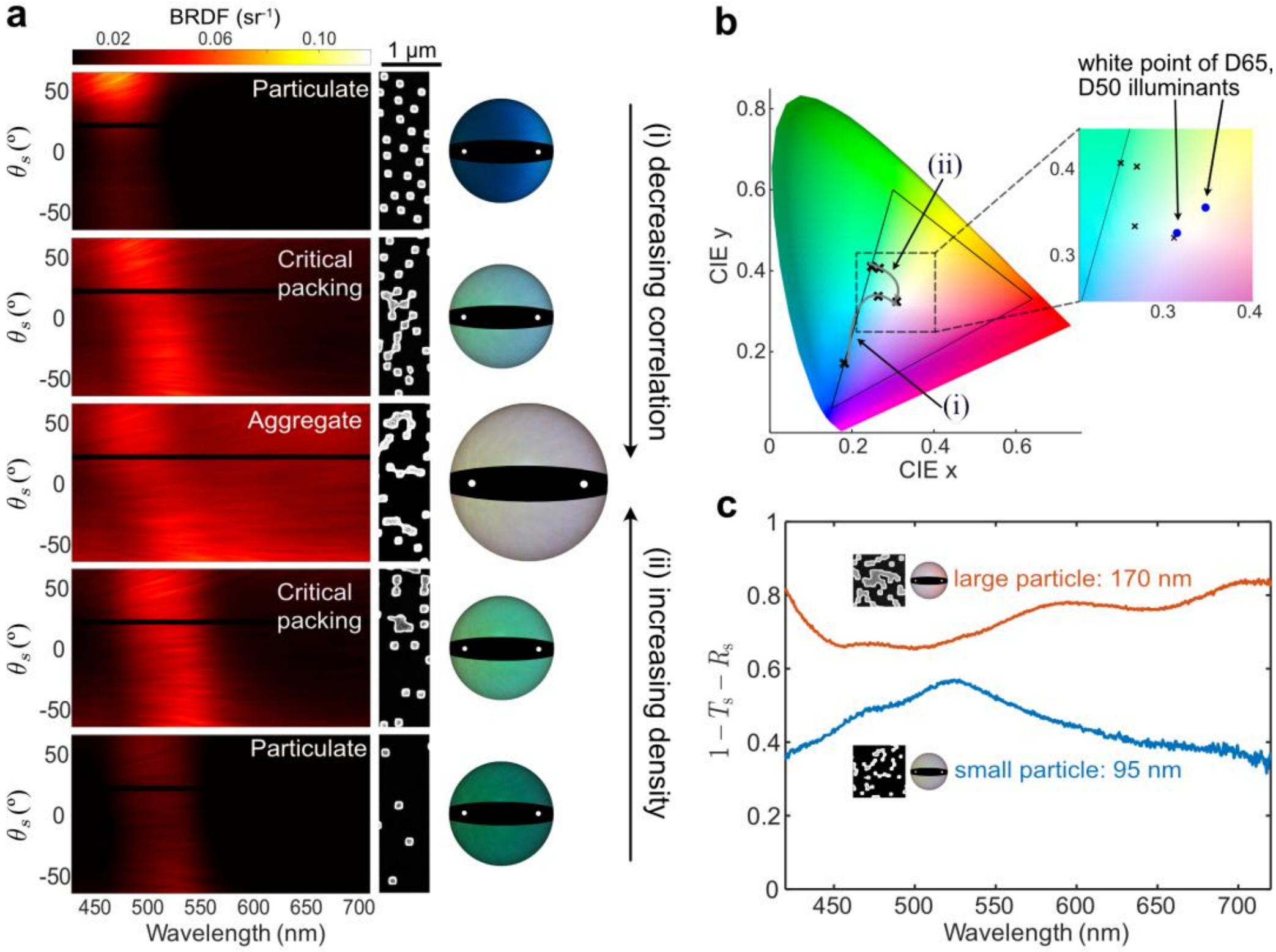

**Figure 5**. **Whitening of diffuse light above critical packing. a** The BRDF and ping-pong ball images are obtained for a fixed metaatom size $\ell = 95$ nm, by increasing the density or decreasing the spatial correlation for an incident angle of 30°. **b** CIE 1931 chromaticity diagram providing the colour coordinates (dark crosses) of the diffuse light in the direction perpendicular to the metasurfaces. The two blue dots indicate the white points for two standard illuminants, D65 and D50. In Suppl. Section S6, we present the same diffuse colours in the CIELAB 3D colour space, where brightness information (L*) is explicitly included. **c** Spectrum of $1 - T_s - R_s$ for the whitest metasurfaces of the set of 27 metasurfaces and for $\theta_i = 10°$. The brightest one shown with a red colour corresponds to $p = 0.1$, $\ell = 170$ nm and $\rho = 10\ \mu\text{m}^{-2}$ in Fig. 2c.

The whitening phenomenon can be intuitively understood through the QNM analysis in Fig. 2. It can be attributed to an inhomogeneous broadening of aggregate topologies, where particles overlap and even cluster. Then numerous modes, which typically cover all visible frequencies, are significantly excited by collimated beams, scattering light at all visible frequencies in all directions.

A key performance metric for white paint or high haze surfaces is the brightness, which quantifies the fraction of the light that is effectively diffused. High brightness values are typically achieved with materials that exhibit strong scattering, often involving complex multilayer structures or thick particulate coatings [21,63]. Figure 5c presents the spectral dependence of the diffuse light – actually, $1 - R_s - T_s$ includes ≈ 15% Si absorption in the blue – for the two metasurfaces that generate the most pronounced whitening. Both metasurfaces have minimal correlation ($p = 0.1$) and maximum density ($\rho = 10\ \mu\text{m}^{-2}$), with the only difference being the metaatom sizes, $\ell = 95$ and 170 nm. The metasurface with the larger metaatom size forms larger aggregates, resulting in a brightness approaching 80% across the entire visible spectrum.

While this value is lower than the brightness achieved by certain animals [64] or ultra-white paints [65], the 145-nm thick metasurfaces strike a relevant balance between brightness and thickness. This makes them a promising choice for compact, photonic CMOS-compatible devices in applications like materials science and lighting, particularly for reducing glare in electronic displays.

## Broadband mirror effect at critical packing

The two preceding effects are observed for diffuse light. Despite their stark contrast in nature, both diffuse and specular lights are influenced by the PDOS and the QNM excitation probabilities. Thus, much like diffuse light, specular light is expected to exhibit rapid variations at critical packing.

This expectation is confirmed in Fig. 6, where we analyze the specular reflection spectra of five metasurfaces transitioning from particulate to aggregate topologies by increasing the silicon post size and disorder correlation. As expected, we first observe a colour transition, from forest green to a dark shade, passing through a distinct scarlet red at critical packing.

Further insight is gained from the reflectivity spectra in Fig. 6b(i). For the particulate metasurface with the smallest metaatoms, the specular reflectivity $R_s$ remains weak across the visible spectrum, peaking below 10%. This metasurface primarily transmits light coherently, with a specular transmission $T_s >$ 80% for $\lambda > 500$ nm. Similarly, the aggregate metasurface with minimal correlation exhibits weak specular reflectivity. Here, the polydispersity of cluster sizes and shapes, combined with their weakly correlated positions, suppresses coherence, leading to dominant diffuse scattering across all wavelengths (Fig. 5).

This dual behaviour explains why aggregates can appear bright white in diffuse reflection while remaining dark in the specular direction. The broad continuum of resonant states ensures efficient scattering of all visible wavelengths into diffuse channels, giving rise to whiteness. At the same time, the randomized phases of these scattered waves suppress specular reflection. The effect is robust across particle sizes large enough to scatter efficiently in the visible, as illustrated in the last three rows of the third column of Fig. 2c.

Between these extremes, the intermediate metasurfaces exhibit significant changes in specular reflectivity as they approach critical packing. The peak reflectivity first increases, reaching 50% in the red, before declining. Notably, the ratio $R_s/T_s$ at critical packing is particularly high, reaching $\approx$ 15 (Fig. 6b(ii)).

A large coherent reflection of 50% is a singular observation. It is neither explained by existing models nor intuitively anticipated by considering the morphology of the critical-packing metasurface in Fig. 6a. To contextualize this observation, we compare the metasurface reflectance with that of a benchmark highly-coherent system: a Si film on glass with the same 145nm thickness. The Si-film reflectance computed using tabulated refractive index data of silicon [45] are shown with dashed curves in Figs. 6b(ii-iii). It exhibits different characteristics, such as an absence of marked peak for $R_s/T_s$ and a reflectance nearly independent of the incident angle. Interestingly, the maximum reflectance of 60% is only slightly higher than that of the metasurface, highlighting the singular capability of the critical-packing metasurface to coherently reflect light despite its strong heterogeneities.

No robust theoretical framework currently exists for modelling metasurfaces near critical packing [21,36, 39,47,60,66]. However, to achieve a qualitative insight into the singular properties of the critical packing metasurface, we attempted to infer effective parameters using effective medium theory [47]. For the two particulate metasurfaces, predictions based on an Extended-Maxwell-Garnett model [67] align well with specular reflectance measurements (Fig. S4), by using the electric and magnetic dipole Mie coefficients of the individual metaatoms. As expected, the model is less quantitative at critical packing, where it significantly underestimates the reflection peak at 670 nm (Fig. S4). While limited confidence can be placed in the model within this regime—where it exceeds its range of validity—we note an increasing role of the magnetic dipole resonance as the metaatom sizes increase to reach critical packing, where we find that the reconstructed permittivities and permeabilities are both remarkably close to zero at the peak wavelength.

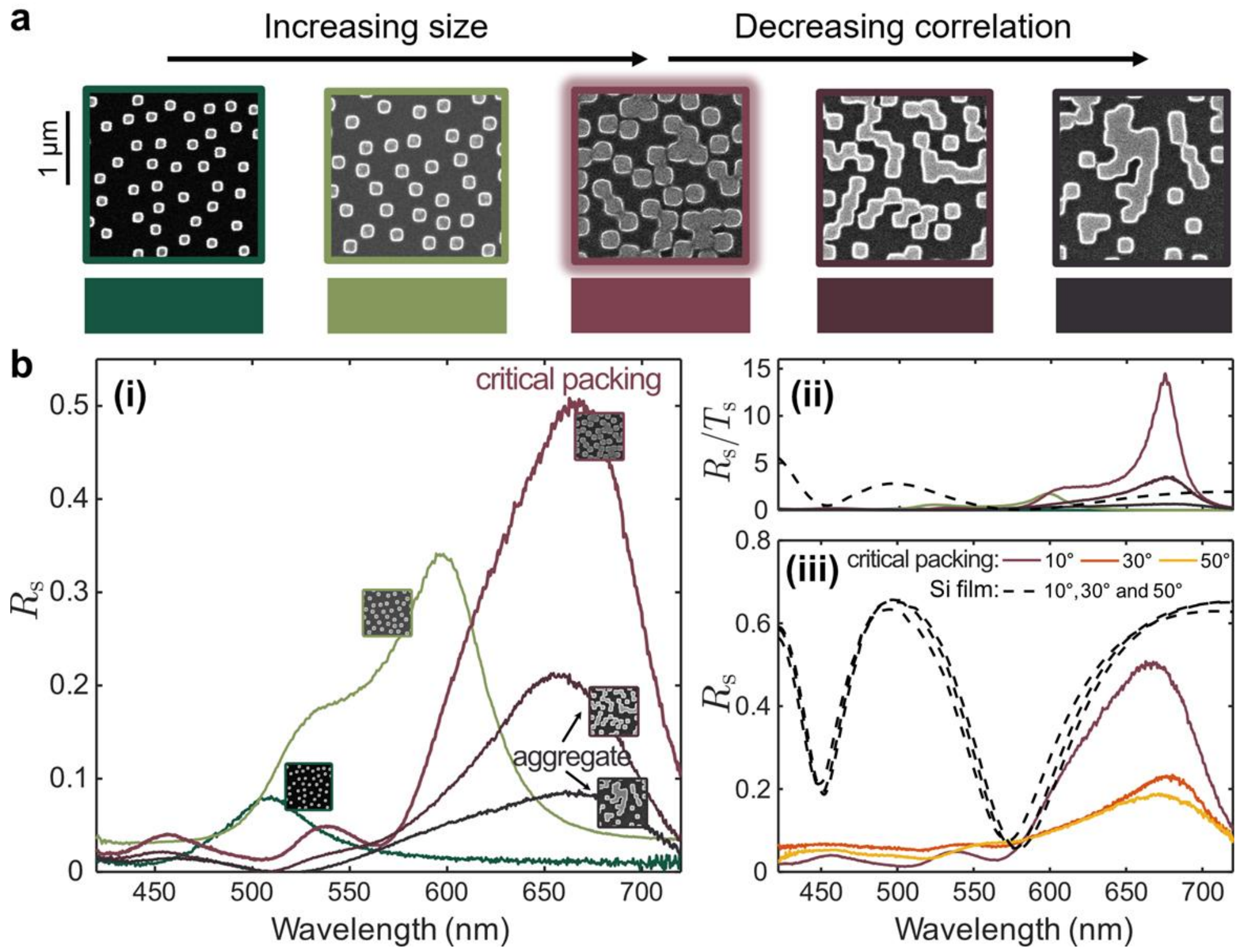

**Figure 6**. **Specular reflectance when transitioning through critical coupling. a** Transitioning through critical packing by increasing the metaatom size of particulate metasurfaces and further reducing the disorder correlation of aggregate metasurfaces. The reflected colour transitions from forest green to a dark shade with a hint of grey; at critical packing, where it exhibits a scarlet red (it is teal green in transmission). **b (i)** Specular reflectance spectra $R_s$ near normal incidence ($\theta_i = 10°$) for the five metasurfaces shown in **a**. **(ii)** $R_s/T_s$ ratio between the specular reflectance and transmission for the three metasurfaces operating around critical packing (the same colours as in **(i)** are used). **(iii)** Specular reflectance at critical packing for different incidences, $\theta_i = 10°$, 30° and 50°. In **(ii)** and **(iii)**, the dashed curves represent computational results for an equivalent 145nm-thick Si film on glass. The nominal metaatom density for all metasurfaces is 10 $\mu m^{-2}$.

The QNM analysis further supports the significance of magnetic effects. At critical packing, the excitation probability distribution for even-symmetry QNMs in Fig. 3a(ii) peaks around the electric dipolar modes of individual metaatoms, while the odd-symmetry QNM PDOS in Fig. S2.5 peaks from magnetic dipolar modes. Since both figures indicate a single dominant QNM with comparable excitation efficiency—such as mode 3 for the even-symmetry QNM in Fig. 3a(ii)—we expect that the metasurface at critical packing exhibits both significant effective permeability and permittivity.

## Outlook

By tuning key technological parameters — such as particle size $\ell$, density $\rho$, and packing fraction $p$ — in silicon metasurfaces, we have investigated the rapid evolution of specular and diffuse optical responses across disordered structures transitioning from particulate to aggregate topologies. A host of interesting visual effects can transpire at the transition, such as a pronounced blue-shift in diffuse colour, a whitening of the diffuse light and a singular mirror effect, which is not intuitively anticipated from a morphology perspective. These results mark a significant advancement in a largely unexplored design space [23].

Our theoretical model of this transition reveals that, at critical packing, the resonant modes undergo a rapid evolution, accompanied by a geometrical phase transition in the photon density of states. The model further highlights that the overlap rate $\tilde{O}$ between metaatoms can serve as an effective unified parameter to track the transition in the full $\rho - p - \ell$ parameter space. It also predicts that the transition occurs within the range $10\% < \tilde{O} < 30\%$.

It is important to distinguish between critical packing and percolation in semi-continuous metal films, both of which involve geometrical phase transitions. In disordered metal metasurfaces, percolation refers to the formation of spanning clusters [68] that connect the entire sample, resulting in a sharp decrease in sheet resistance. These nearly continuous systems behave similarly to homogeneous films, dominated by extended networks of hot spots, which play a key role in absorption and nonlinear optics [39,40].

In contrast, the critical-packing phase transition examined here affects local connectivity, where only a small number of metaatoms are interconnected, forming small clusters. This occurs much earlier than percolation, long before a continuous film is established. The morphology remains particulate or aggregate, and the abrupt shift in optical behaviour arises from scattering, mode hybridization, and diffusion rather than percolative transport. Thus, while this transition is reminiscent of percolation due to its geometrical phase transition, critical packing represents a distinct regime that broadens the design space for disordered metasurfaces.

The use of high-index-contrast Si subwavelength metaatoms with a few distinct low-loss resonances is central in the present work. By comparison, low-index-contrast metaatoms, such as those fabricated from polymers, are expected to exhibit much weaker optical effects. Metallic metaatoms, on the other hand, behave quite differently: at critical packing, they support numerous additional resonances with intense fields localized in the nanogaps between nearly touching particles. This produces a photonic density of states (PDoS) landscape distinct from that shown in Fig. 3a. Because these extra modes are lossy and span a broad spectral range, absorption strongly influences the critical-packing transition, diminishing the visibility of certain effects—such as the blue shift—that we observe in dielectric metasurfaces. However, these reductions may be offset by the striking emergence of metallic cluster.

Similarly, variations in the morphology of the metaatoms—such as changes in size or shape—do influence their resonance and thus the resulting colour. However, these variations do not hinder the observation of the transition phenomena. In particular, the pronounced dispersity of appearances in Fig. 2 is expected to remain robust against morphological fluctuations. A noteworthy special case arises with elongated particles, especially when their orientation is not random. In such situations, anisotropy is preserved within the metasurfaces, and new polarization-dependent effects may emerge in both diffuse and specular light.

A deeper statistical characterization of the QNMs, including spectral and spatial fluctuations of the PDoS, would represent an exciting avenue for future investigation. Such an analysis could reveal additional signatures of the transition and provide complementary insights beyond the mean statistics considered here [69].

We expect that metasurfaces engineered at critical packing will serve as platforms for a broad spectrum of applications, spanning both fundamental research and technological implementation, by enabling precise control over light–matter interactions. Moreover, their compatibility with scalable fabrication methods—such as soft lithography [70] and bottom-up colloidal lithography [71,72]—and their robustness to imperfections make them promising for large-area production.

## References

1. M. Imada, A. Fujimori, Y. Tokura, "Metal-insulator transitions", Rev. Mod. Phys. **70**, 1039 (1998).
2. E. Yablonovitch, "Inhibited spontaneous emission in solid-state physics and electronics", Phys. Rev. Lett. **58**, 2059 (1987).
3. E. A. Cornell, C. E. Wieman, "Nobel Lecture: Bose-Einstein condensation in a dilute gas, the first 70 years and some recent experiments", Rev. Mod. Phys. **74**, 875 (2002).
4. P. Lalanne, W. Yan, V. Kevin, C. Sauvan, and J.-P. Hugonin, "Light interaction with photonic and plasmonic resonances", Laser & Photonics Reviews **12**, 1700113 (2018).

5. A. Chabanov, M. Stoytchev, A. Genack, "Statistical signatures of photon localization", Nature **404**, 850–853 (2000).
6. M. Davy, Z. Shi, J. Wang, X. Cheng, and A. Z. Genack, "Transmission Eigenchannels and the Densities of States of Random Media", Phys. Rev. Lett. **114**, 033901 (2015)
7. M. Davy and A. Z. Genack, "Selectively exciting quasi-normal modes in open disordered systems", Nat. Commun. **9**, 4714 (2018).
8. M. Balasubrahmaniyam, S. Mondal and S. Mujumdar, "Necklace-State-Mediated Anomalous Enhancement of Transport in Anderson-Localized non-Hermitian Hybrid Systems", Phys. Rev. Lett. **124**, 123901 (2020).
9. L. Chen, S. M. Anlage and Y. V. Fyodorov, "Statistics of Complex Wigner Time Delays as a Counter of S-Matrix Poles: Theory and Experiment", Phys. Rev. Lett. **127**, 204101 (2021).
10. R. Monsarrat, R. Pierrat, A. Tourin and A. Goetschy, "Pseudogap and Anderson localization of light in correlated disordered media", Phys. Rev. Res. **4**, 033246 (2022).
11. A. Goetschy and S.E. Skipetrov, "Euclidean random matrices and their applications in physics", arXiv preprint, arXiv:1303.2880.
12. D. J. Thouless, "Maximum metallic resistance in thin wires", Phys. Rev. Lett. **39**, 1167 (1977).
13. S. E. Skipetrov, I. M. Sokolov, "Absence of Anderson Localization of Light in a Random Ensemble of Point Scatterers", Phys. Rev. Lett. **112**, 023905 (2014).
14. A. Z. Genack and S. Zhang, "Interference and Modes in Random Media", in *Tutorials in complex photonic media,* Noginov, M.A., Dewar, G., McCall, M.W. and Zheludev, N.I. eds. (Bellingham, US. SPIE, 2009).
15. L.S. Froufe-Pérez, M. Engel, J.J. Sáenz, and F. Scheffold, "Band gap formation and Anderson localization in disordered photonic materials with structural correlations", Proc. Natl. Acad. Sci. U.S.A. **114**, 9570-9574 (2017).
16. G. J. Aubry, L. S. Froufe-Pérez, U. Kuhl, O. Legrand, F. Scheffold, F. Mortessagne, "Experimental tuning of transport regimes in hyperuniform disordered photonic materials", Phys. Rev. Lett. **125**, 127402 (2020).
17. H. Cao, "Review on latest developments in random lasers with coherent feedback", J. Phys. A: Math. Gen. **38**,10497 (2005).
18. S. A. Schulz, R. F. Oulton, M. Kenney et al., "Roadmap on photonic metasurfaces", Appl. Phys. Lett. **124**, 260701 (2024).
19. A. I. Kuznetsov, M. L. Brongersma, J. Yao, et al., "Roadmap for optical metasurfaces", ACS photonics **11**, 816-865 (2024).
20. S. Yu, C.-W. Qiu, Y. Chong, S. Torquato, N. Park, "Engineered disorder in photonics", Nature Reviews **6**, 226-243 (2021).
21. K. Vynck, R. Pierrat, R. Carminati, L. S. Froufe-Pérez, F. Scheffold, R. Sapienza, S. Vignolini, and J. J. Sáenz, "Light in correlated disordered media", Rev. Mod. Phys. **95**, 045003 (2023).
22. Z. Hu, C. Liu and G. Li, "Disordered optical metasurfaces: from light manipulation to energy harvesting", Advances in Physics: X **8**, 2234136 (2023).
23. P. Lalanne, M. Chen, C. Rockstuhl, A. Sprafke, A. Dmitriev, K. Vynck, "Disordered optical metasurfaces: basics, design and applications", Adv. Opt. Photon. **17**, 45-112 (2025).
24. Z. Yu, H. Li, W. Zhao, P. S. Huang, Y. T. Lin, J. Yao, W. Li, Q. Zhao, P. C. Wu, B. Li, P. Genevet, Q. Song, P. Lai, "High-security learning-based optical encryption assisted by disordered metasurface", Nat. Commun. **15**, 2607 (2024).
25. M. Song, D. Wang, Z. Kudyshev, Y. Xuan, Z. Wang, A. Boltasseva, V. Shalaev, A. Kildishev, "Enabling Optical Steganography, Data Storage, and Encryption with Plasmonic Colours", Laser & Photonics Reviews **15**, 2000343 (2021).
26. M. A. Fusella, R. Saramak, R. Bushati et al., "Plasmonic enhancement of stability and brightness in organic light-emitting devices", Nature **585**, 379-382 (2020).
27. P. Mao, C. Liu, X. Li, M. Liu, Q. Chen, M. Han, S. A. Maier, E. H. Sargent, S. Zhang, "Single-step-fabricated disordered metasurfaces for enhanced light extraction from LEDs", Light: Science & Applications **10**, 180 (2021).

28. H. Chu, X. Xiong, N. X. Fang, F. Wu, R. Jia, R. Peng, M. Wang, Y. Lai, "Matte surfaces with broadband transparency enabled by highly asymmetric diffusion of white light", Sci. Adv. **10**, eadm8061 (2024).
29. V. A. Maiorov, "Window Glasses: State and Prospects", Optics and Spectroscopy **124**, 594-608 (2018).
30. P. Cencillo-Abad, D. Franklin, P. Mastranzo-Ortega, J. Sanchez-Mondragon, D. Chanda, "Ultralight plasmonic structural colour paint," Sci. Adv. **9**, eadf7207 (2023).
31. Y. Jiang, D. Chen, Z. Zhang, X. Wu, Y. Tu, Z. Zheng, L. Mao, W. Li, Y. Ma, J. Du, W.-J. Wang, P. Liu, "Meta-Structured Covalent Organic Frameworks Nano-Coatings with Active and Angle-Independent Structural Colouration", Adv. Mater. **36**, 2311784 (2024).
32. P. Mao, C. Liu, Y. Niu, Y. Qin, F. Song, M. Han, R. E. Palmer, S. A. Maier, S. Zhang, "Disorder-Induced Material-Insensitive Optical Response in Plasmonic Nanostructures: Vibrant Structural Colours from Noble Metals", Adv. Mater. **33**, 2007623 (2021).
33. P. Mao, C. Liu, F. Song, M. Han, S. A. Maier, S. Zhang, "Manipulating disordered plasmonic systems by external cavity with transition from broadband absorption to reconfigurable reflection", Nat. Commun. **11**, 1538 (2020).
34. K. Vynck, R. Pacanowski, A. Agreda, A. Dufay, X. Granier, P. Lalanne, "The visual appearances of disordered optical metasurfaces", Nat. Mater. **21**, 1035-41 (2022).
35. D. Arslan, A. Rahimzadegan, S. Fasold, M. Falkner, W. Zhou, M. Kroychuk, C. Rockstuhl, T. Pertsch, I. Staude, "Toward perfect optical diffusers: dielectric huygens metasurfaces with critical positional disorder", Adv. Mater. **34**, 2105868 (2022).
36. S. Torquato and J. Kim, "Nonlocal Effective Electromagnetic Wave Characteristics of Composite Media: Beyond the Quasistatic Regime", Phys. Rev. X **11**, 021002 (2021).
37. A. Agreda, T. Wu, A. Hereu, M. Treguer-Delapierre, G. L. Drisko, K. Vynck, P. Lalanne, "Tailoring iridescent visual appearance with disordered resonant metasurfaces", ACS Nano **17**, 6362-6372 (2023).
38. Y. Billiet, V. Chevalier, S. Kostcheev, H. Kadiri, S. Blaize, A. Rumyantseva, G. Lérondel, "Engineering of Diffuse Structural Colours", Adv. Optical Mater. **12**, 2302165 (2024).
39. V. M. Shalaev, "Electromagnetic properties of small-particle composites", Phys. Rep. **272**, 61-137 (1996).
40. R. Ron, E. Haleva, A. Salomon, "Nanoporous Metallic Networks: Fabrication, Optical Properties, and Applications", Adv. Mater. **30**, 1706755 (2018).
41. M. Patel, "Poisson-disc-sampling: Matlab script for n-dimensional Poisson-disc sampling", *GitHub* https://github.com/mohakpatel/Poisson-Disc-Sampling (2016).
42. F. E. Nicodemus, J. C. Richmond, J. J. Hsia, I. W. Ginsberg, and T. Limperis, "Geometric Considerations and Nomenclature for Reflectance", National Bureau of Standards (1977).
43. A. Lagendijk, B. A. van Tiggelen, "Resonant multiple scattering of light", Phys. Rep. **270**, 143-215 (1996).
44. T. Wu, D. Arrivault, W. Yan, P. Lalanne, "Modal analysis of electromagnetic resonators: user guide for the MAN program", Comput. Phys. Commun. **284**, 108627 (2023).
45. M. A. Green, M. J. Keevers, "Optical properties of intrinsic silicon at 300 K", Prog. Photovolt.: Res. Appl. **3**, 189-192 (1995).
46. M. J. Huttunen, K. Dolgaleva, P. Törmä, and R. W. Boyd, "Ultra-strong polarization dependence of surface lattice resonances with out-of-plane plasmon oscillations", Opt. Express **24**, 28279 (2016).
47. H. Benisty, J. J. Greffet and P. Lalanne, *Introduction to Nanophotonics* (Oxford University Press, Oxford, 2022).
48. G. Gallavotti, *Statistical Mechanics: A Short Treatise* (Springer, 1999).
49. C. Liu, A. Di Falco, D. Molinari al., "Enhanced energy storage in chaotic optical resonators", Nat. Photon. **7**, 473-478 (2013).
50. R. Albert, A. L. Barabási, "Statistical mechanics of complex networks", Rev. Mod. Phys. **74**, 47 (2002).
51. N. J. Halas, S. Lal, W.-S. Chang, S. Link, and P. Nordlander, "Plasmons in Strongly Coupled Metallic Nanostructures", Chemical Reviews **111**, 3913-3961 (2011).

52. A. I. Kuznetsov, A. E. Miroshnichenko, M. L. Brongersma, Y. S. Kivshar, B. Luk'yanchuk, "Optically resonant dielectric nanostructures", Science **354**, aag2472 (2016).
53. W. Yang, S. Xiao, Q. Song, et al., "All-dielectric metasurface for high-performance structural color", Nat. Commun. **11**, 1864, 2020.
54. D. Wang, Z. Liu, H. Wang, M. Li, L. Jay Guo and Cheng Zhang, "Structural color generation: from layered thin films to optical metasurfaces", Nanophotonics **12**, 1019-1081 (2023).
55. Parker R. Wray and Harry A. Atwater, "Light–Matter Interactions in Films of Randomly Distributed Unidirectionally Scattering Dielectric Nanoparticles", ACS Photonics **7**, 2105-2114 (2020).
56. T. J. Antosiewicz, S. P. Apell, M. Zäch, I. Zori, and C. Langhammer, "Oscillatory Optical Response of an Amorphous Two-Dimensional Array of Gold Nanoparticles", Phys. Rev. Lett. **109**, 247401 (2012).
57. J. Pellegrino, R. Bourgain, S. Jennewein, Y. R. P. Sortais, A. Browaeys, S. D. Jenkins, and J. Ruostekoski, "Observation of Suppression of Light Scattering Induced by Dipole-Dipole Interactions in a Cold-Atom Ensemble", Phys. Rev. Lett. **113**, 133602 (2014).
58. M. O. Araújo, W. Guerin and R. Kaiser, "Decay dynamics in the coupled-dipole model", J. Mod. Opt. **65**, 1345-1354 (2018).
59. G. Voronovich, *Wave Scattering from Rough Surfaces* (Springer Science & Business Media, 2013).
60. B. X. Wang, C. Y. Zhao, "The dependent scattering effect on radiative properties of micro/nanoscale discrete disordered media", Annual Reviews of Heat Transfer **23**, 231-353 (2020).
61. D. W. Watson, S. D. Jenkins, and J. Ruostekoski, "Point dipole and quadrupole scattering approximation to collectively responding resonator systems", Phys. Rev. B **96**, 035403 (2017).
62. A. Reyes-Coronado, G. Morales-Luna, O. Vázquez-Estrada, A. García-Valenzuela, and R. G. Barrera, "Analytical modeling of optical reflectivity of random plasmonic nano-monolayers", Opt. Express **26**, 12660 (2018).
63. D. S. Wiersma, "Disordered photonics", Nat. Photonics **7**, 188-196 (2013).
64. M. Burresi, L. Cortese, L. Pattelli, M. Kolle, P. Vukusic, D.S. Wiersma, U. Steiner, and S. Vignolini, "Bright-white beetle scales optimise multiple scattering of light", Sci. Rep. **4,** 6075 (2014).
65. X. Li, J. Peoples, P. Y. Yao, and X. Ruan, "Ultra-White BaSO4 Paint and Film with Remarkable Radiative Cooling Performance", ACS Appl. Mater. Interfaces **13**, 21733-739 (2021).
66. Z. M. Sherman, D. J. Milliron, and T. M. Trusket, "Distribution of Single-Particle Resonances Determines the Plasmonic Response of Disordered Nanoparticle Ensembles", ACS Nano **18**, 21347-21363 (2024).
67. V. A. Markel, "Introduction to the Maxwell Garnett approximation: tutorial", J. Opt. Soc. Am. A **33**, 1244-1256 (2016).
68. D. Stauffer and A. Aharony, *Introduction to Percolation Theory* (Taylor & Francis, 1994).
69. V. Krachmalnicoff, E. Castanié, Y. De Wilde, R. Carminati, "Fluctuations of the Local Density of States Probe Localized Surface Plasmons on Disordered Metal Films", Phys. Rev. Lett. **105**, 183901 (2010).
70. D. Qin, Y. Xia, G. M. Whitesides, "Soft lithography for micro-and nanoscale patterning", Nature Protocols **5**, 491-502 (2010).
71. V. Su, C. H. Chu, G. Sun, D. P. Tsai, "Advances in optical metasurfaces: fabrication and applications", Opt. Express **26**, 13148-82 (2018).
72. G. Cossio, R. Barbosa, B. Korgel, E. T. Yu, "Massively Scalable Self-Assembly of Nano and Microparticle Monolayers via Aerosol Assisted Deposition", Adv. Mater. **36**, 2309775 (2024).
73. A. Gras, W. Yan, P. Lalanne, "Quasinormal-mode analysis of grating spectra at fixed incidence angles", Opt. Lett. **44**, 3494-3497 (2019).

## Acknowledgements

PL acknowledges financial support from the Grand Research Program LIGHT Idex of Bordeaux University and the European Research Council Advanced grant (Project UNSEEN No. 101097856). The authors thanks Louise-Eugénie Bataille, Arnaud Tizon, Philippe Teulat and Louis Bellando for their help in developing the goniospectrometer setup. They acknowledge fruitful interactions with Romain Pacanowski, Xavier Granier, Pascal Barla, Glenna Drisko, Mona Treguet-Delapierre and Thomas

Christopoulos. Sample fabrication was supported by LAAS-CNRS micro and nanotechnologies platform, a member of the RENATECH French national network.

## Author contributions

PL and KV conceived and planned the project. PL and AA designed the metasurfaces. FC fabricated them. AA and MC performed the experimental measurements and calibrated photographs. MC performed the electromagnetic analysis with the help of TW. MC and PL wrote the manuscript. All authors discussed the results and their interpretation.

## Competing interests

The authors declare no competing interests.

## Methods

**Sample fabrication.** The metasurfaces were fabricated by etching structures in a 150nm-thick polycrystalline silicon (Si-poly) layer. First, Si-poly was deposited on both sides of a 4-inch fused silica (FS) wafer by LPCVD at 605°C. Then, Si-poly was removed from one face of the FS wafer using fluorinated based reactive ion etching (F-RIE). Sample was then covered by a 160nm-thick layer of a negative resist (maN2405 – Micro Resist Technology) followed by a 40nm-thick conductive layer (ELECTRA92 – AllResist) to avoid charging effects during subsequent electron beam lithography.

Samples were exposed by EBL with a beam energy of 20keV at a current of 135pA with a stepsize of 4nm and a nominal dose of 150µC/cm$^2$. Resist was developed using MF-CD-26 (Micropposit) at 20°C during 50s, rinsed with deionized water for 1 minute and dried with nitrogen. Negative maN resist patterns were transferred in Si-poly layer using F-RIE and remaining resist mask was removed by oxygen plasma.

Depending on density and applied dose, one can vary the size of the squares in the resist after development using proximity effects. Increasing the dose can then lead to fusion of squares as it can be seen in figures 1, 2, 5 and 6.

**Visual observation with the ping-pong ball.** A $200\ \mu\mathrm{m}$ supercontinuum laser beam (Leukos, Rock 400) is slightly focused onto the metasurface samples. A semi-translucent half-ping-pong ball is positioned at the centre of the metasurface, serving as a hemispherical screen for the diffusely reflected light. The images of the ping-pong ball are recorded with a Canon EOS 1000D camera mounted on an optical table with fixed exposure times for each series. Full details on the characterization of diffuse and specular light, including the diffusive transmission properties of the ping-pong ball, are provided in Section S5.

**BRDF Characterization.** BRDF measurements are conducted using a custom-built goniospectrophotometric setup equipped with a supercontinuum laser and a 1 mm-diameter optical fiber connected to a spectrometer (Ocean Insight, HDX). The incidence $(\theta_i)$ and scattering $(\theta_s, \phi_s)$ angles are precisely controlled by three stepper motor rotation stages (Newport URS75, URS150, and SR50CC).

To ensure measurements in the visible range, the unpolarized supercontinuum laser beam is filtered using a short-pass filter (Schott KG-1). The incident laser radiant flux is measured within the same setup, with the fiber detector aligned to capture the focused laser beam.

**Electromagnetic computational results.** The QNM are computed for a null in-plane wavevector with the QNMEig solver of the freeware MAN [44]. This solver automatically identifies a large number of modes near a user-defined complex frequency. The computed eigenvectors include both QNMs and numerical PML modes. To eliminate PML modes, we implement a systematic procedure detailed in Supp. Section S2.1. After filtering, we normalize the QNMs [44,73] and rank them based on their excitation coefficient [44]

$$|\alpha_m| \sim |2(\varepsilon_{Si} - \varepsilon_b) Q_m \iiint \mathbf{E}_b \cdot \tilde{\mathbf{E}}_m \, d^3r, \quad (2)$$

for a normally incident plane wave at a frequency matching the real part of the QNM frequency. Here, $\varepsilon_b = 1$ and $\varepsilon_{Si} = 20 - 0.5i$ denote the background and silicon relative permittivities, respectively. The negative sign of the imaginary part reflects COMSOL's use of the $e^{i\omega t}$ time-harmonic convention, where a negative imaginary component corresponds to material loss. The spatial overlap integral between the incident electric field $\mathbf{E}_b$ and the normalized QNM electric field $\tilde{\mathbf{E}}_m$ is evaluated within the silicon posts of the supercell.

Figure 3a using even-symmetric QNMs collected from 40 independent metasurface realizations. 20 of these 40 realizations are displayed in Figs. S2.2-S2.4, where each dot represents a QNM complex frequency, with the colour encoding the excitation coefficients (blue for low values, red for high). In each realization, only a small subset of QNMs is strongly excited. On average, approximately 190 QNMs are computed per realization, resulting in a total of around 7,600 QNMs used to estimate the photonic density of states (PDoS) maps. Additional realizations for odd-symmetric modes at critical packing density are presented in Fig. S2.5, with the corresponding PDoS map shown in Fig. S2.6.

Importantly, note that consistent results are observed for supercells four times larger (Figs. S2.7-S2.8), reinforcing our confidence that the statistical data from the 1.2×1.2 µm² supercells are sufficiently converged to support our conclusions on the transition of the PDoS across particulate, critical-packing, and aggregate topologies.

The COMSOL model for computing the QNMs will be provided the freeware MAN V9 and following versions.

# Supplementary Information for "Emergent scattering regimes in disordered metasurfaces near critical packing"

M. Chen[1], A. Agreda[1‡], T. Wu[1], F. Carcenac[2], K. Vynck[3] and P. Lalanne[1*]

[1] LP2N, CNRS, IOGS, Université Bordeaux, Talence, France

[2] LAAS-CNRS, Université de Toulouse, CNRS, Toulouse, France

[3] Institut Lumière Matière, CNRS, Université Claude Bernard Lyon 1, Villeurbanne, France

[‡]Current address: ELORPrintTec, Allée Geoffroy Saint-Hilaire, F-33600 Pessac, France

* philippe.lalanne@institutoptique.fr

## S1. Particulate disordered metasurfaces with increasing nanopost sizes

This section examines the visual appearance of disordered metasurfaces with particulate topologies.

In addition to the 27 metasurfaces shown in Fig. 2, we also fabricated particulate metasurfaces, systematically adjusting the nanopost square size from 95 to 165 nm in 15 nm increments. The metasurfaces consist of silicon nanoposts of varying sizes (with a fixed height of 145 nm) on a glass substrate. Two packing fractions are considered: $p = 0.1$ and 0.5.

Figure S1a presents the specular (rectangle boxes) and diffuse colours (ping-pong ball images) observed for metasurfaces with $p = 0.1$ and $\rho = 3.14\ \mu\text{m}^{-2}$, illuminated at a 45° incidence angle. The ping-pong ball images cover a broader colour spectrum compared to Fig. 2, with colours transitioning smoothly from navy blue to green, and then to gold and caramel as particle size increases. Saturated red is absent because the largest nanoparticles support multiple resonance modes in the visible spectrum.

Figure S1b presents the specular and diffuse colour distributions for metasurfaces with an increased packing fraction of 0.5. This transition transforms a single, angularly invariant diffuse colour into an iridescent effect. In this regime, diffuse light is suppressed near the specular direction and sharply enhanced in a ring-shaped pattern at larger angles. This suppression is observed across all samples as the resonance wavelength increases. Moreover, the angular position of the intense diffuse halo shifts with changing colour, with the relative angle between the specular reflection and the halo increasing for longer wavelengths.

The role of structural correlations in diffuse reflectance can be understood by analysing the structure factor $S_r$ of the disordered pattern, as shown in Fig. S1d(i) [1,2]. For the weakly correlated sample ($p = 0.1$), $S_r \approx 1$ for all angles and wavelengths, indicating the absence of constructive or destructive interference. However, increasing the packing fraction $p$ induces strong short-range order, leading to a suppression of the structure factor $S_r$ at low momentum transfers and the formation of a prominent peak. At $p = 0.5$, the structure factor modifies the scattering pattern of individual particles, leading to diffuse intensity suppression around the specular reflection direction. This diffuse iridescence becomes even more evident when varying both the incident and viewing angles. Figure S1c illustrates this effect using ping-pong ball images of short-range correlated metasurface (size 150 nm, $\rho = 2\ \mu\text{m}^{-2}$) under different illumination angles. As the incident angle changes, the dark and bright regions shift, following the specular reflection direction. This analysis now provides a direct, quantitative link between the engineered correlations and the resulting optical response: the suppression of $S_r$ causes the quenching of diffuse light near the specular direction, while the dispersive peak in $S_r$ gives rise to the bright, coloured halo.

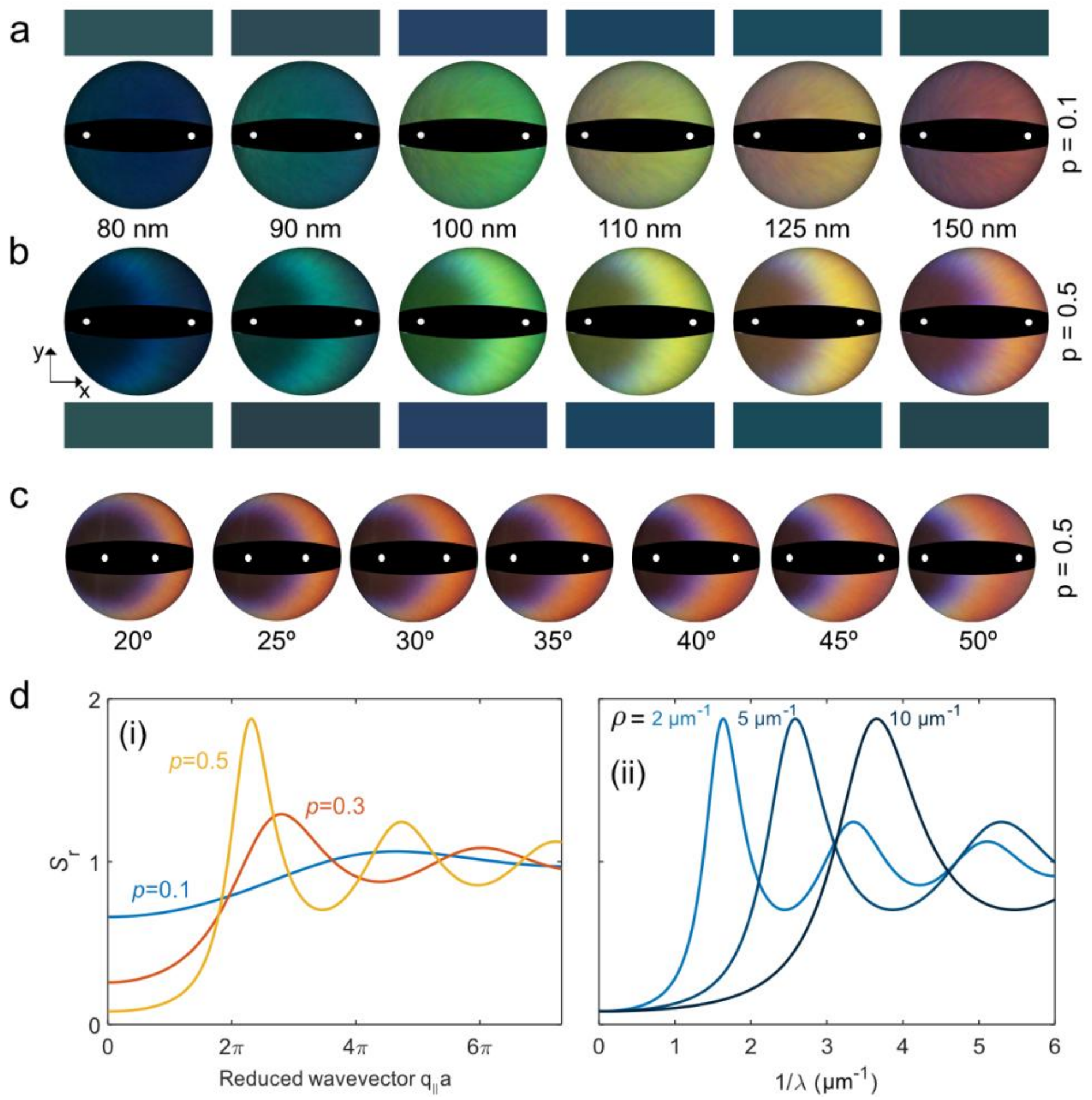

**FIG. S1** The iridescence of correlated disorder: Broadband Quenching. **a** Ping-pong ball images of the diffuse light (bottom) and specular colours (top) of randomly disordered metasurfaces ($\rho = 3.14\ \mu m^{-2}; p = 0.1$) for increasing particle sizes, 80, 90, 100 nm, etc. Metasurfaces are illuminated at 45° incidence. **b** Same for short-range correlated metasurfaces with $p = 0.5$. **c** Diffuse colour of the short-range correlated metasurface of size 150 nm ($\rho = 2\ \mu m^{-2}; p = 0.5$) under varying incident angle illumination. The white dots indicate the incident (rightmost dots) and specular (leftmost dots) directions. **d** Structure factor characterization. (i) static structure factor $S_r$ as a function of the reduced in-plane wavevector $q_{||}a$ for three different packing fractions $p$, where $a = 1/\sqrt{\rho}$ is the mean interparticle distance. (ii) $S_r$ plotted as a function of inverse wavelength for a fixed packing fraction of $p = 0.5$ and for three different particle densities $\rho$.

Figure S1d(ii) shows that as the particle density $\rho$ increases, the primary peak of the structure factor and its associated depleted zone shift to shorter wavelengths (larger $1/\lambda$). This directly corresponds to a widening of the quenching cone for diffuse light at higher densities and explains the dispersive nature of this quenching, where longer wavelengths are suppressed over a larger angular range.

Unlike the strong variations in diffuse colour with metaatom size, the specular colour remains relatively constant, as shown in the lower row of Fig. S1b and the upper row of Fig. S1a. In fact, the overall effect of particle size and correlation on specular light is minor due to the low nanoparticle density.

The visual appearance of an object results from a subtle interplay between its specular and diffuse colours. For the correlated disordered metasurface in Fig. S1b, the appearance is predominantly specular in regions where diffuse scattering is suppressed. Conversely, at larger observation angles, the appearance is dominated by diffuse colours, which are entirely different from the specular reflection.

## S2. Statistical analysis of the QNM computational data

### S2.1 Sorting QNMs and removal of PML modes

This section outlines a systematic approach for distinguishing QNMs from PML modes.

Using **MAN** solver [3], we first compute two eigenspectra of the same metasurface for two different PML implementations, labeled 'PML 1' and 'PML 2' (Fig. S2.1). The PMLs have similarly performance, differing only in their absorption wavelengths—450 nm for 'PML 1' and 600 nm for 'PML 2'. Then we compute the QNM electric "energy" $\xi_m = \iiint \tilde{\mathbf{D}}_m \cdot \tilde{\mathbf{E}}_m \, d^3r$, where the integration volume encompasses the monolayer of the metasurface (all silicon nanoposts and air between them) in the supercell. The electric "energy" is further normalized to its maximum value, with red and blue colours indicating its relative strength (see the two colorbars).

The removal of PML modes follows three key steps:

1. QNMs are largely insensitive to changes in PML parameters, whereas PML modes exhibit strong dependence. QNMs are identified by selecting eigenfrequencies that are nearly identical for the two PMLs, with a maximum relative variation of 3% in our study. In the rightmost zooming window, one can see that the red vertical lines representing computation with PML 1 and the blue horizontal lines representing computation with PML 2 superimpose to form crosses, indicating QNMs. This superimposition is not observed for the PML modes in the leftmost zooming window.

2. Ambiguity may remain. For instance, two PML modes from different PML configurations may accidentally exhibit nearly identical eigenfrequencies. If the two values of $\xi_m$ are both smaller than 0.5, we eliminate the two eigenvectors from the data set. This is because modes with weak electric "energy" are likely PML modes, whereas those with high electric "energy" are classified as QNMs.

3. Finally, when the modes have both their values of $\xi_m$ larger than 0.5 (this happens rarely), we examine their mode profiles. Significant variations in the mode profiles across PML configurations provide further confirmation that these modes are PML artifacts rather than true QNMs.

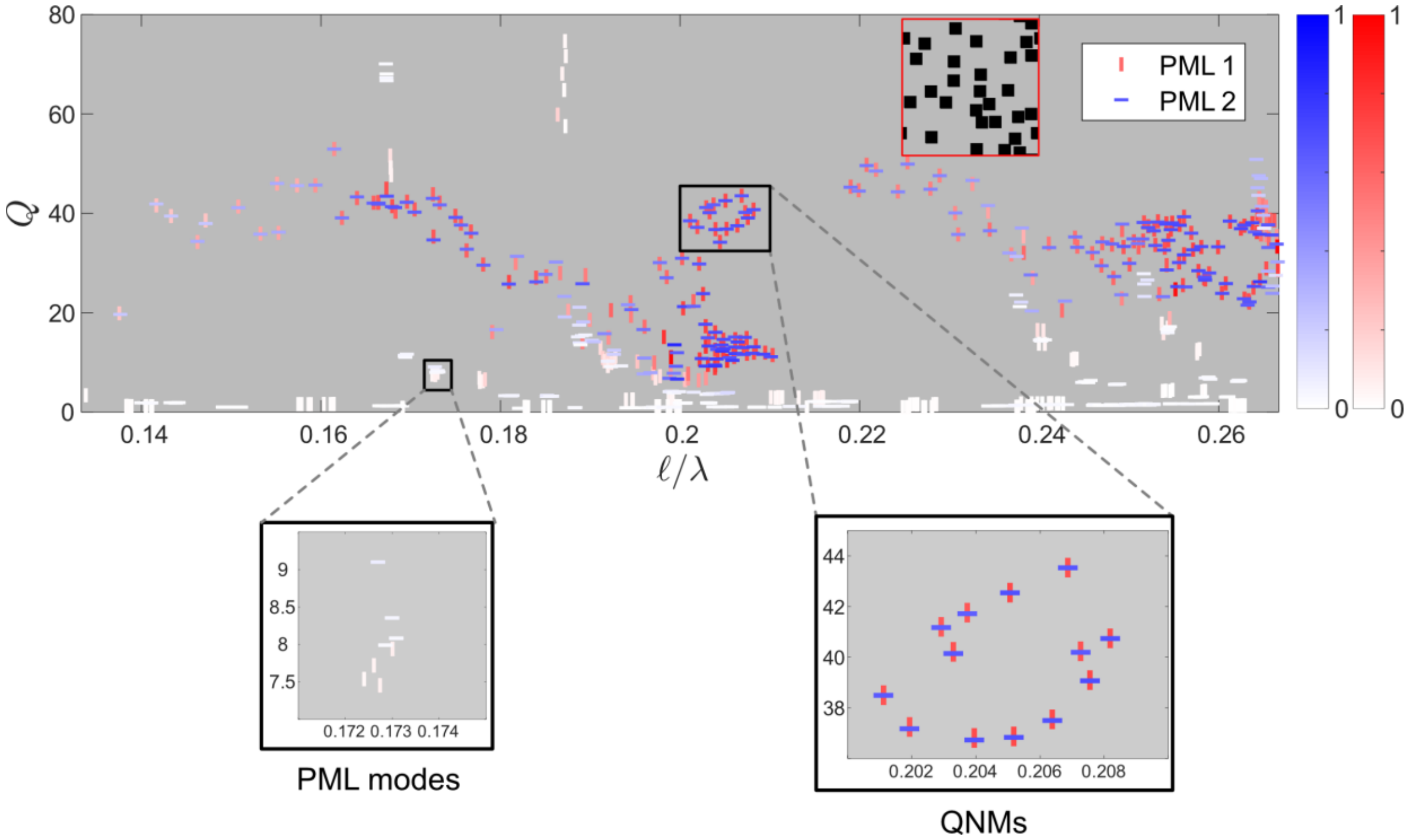

**FIG. S2.1** Discrimination between QNMs and PML-modes for one realization of critical packing disordered metasurface. The material and geometrical parameters are the same as in Fig. 3a(ii). The complex frequencies of even-symmetric modes are computed using two PMLs that have different parameters. They are displayed with red dots and blue crosses, respectively. Light blue or light red marks indicate modes for which the normalized electric 'energy' integrated over all silicon nanoposts is small.

### S2.2 Row data for 20 independent disorder realizations: even symmetry QNMs

The configurational averaging in Fig. 3, used to compute the electromagnetic DOS, is based on 40 independent disorder realizations. Figures S2.2–S2.4 present complex-frequency maps for 20 of these realizations, focusing on even-symmetric QNMs. The remaining 20 realizations are omitted for brevity. A key observation from these individual maps is that the variation between different realizations is relatively small, indicating weak dispersion across disorder configurations.

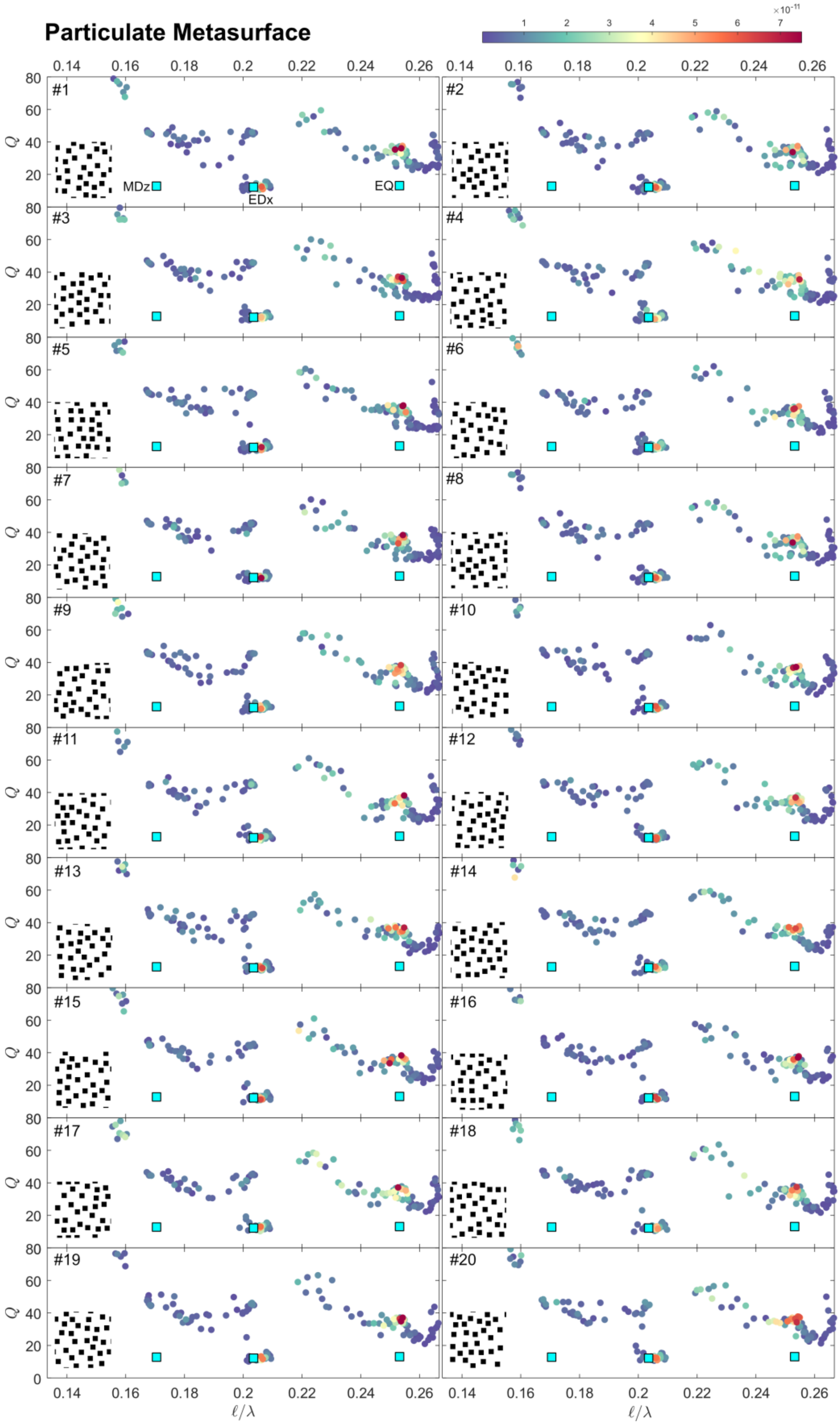

**FIG. S2.2** Complex eigenfrequency maps of particulate metasurfaces (packing fraction 0.5) of 20 independent disorder realizations for even-symmetric QNMs. The parameters of all metasurfaces are the same as in Fig. 3a(i). The insets at left bottom of each map show the distribution of meta-atoms inside the unit cell.

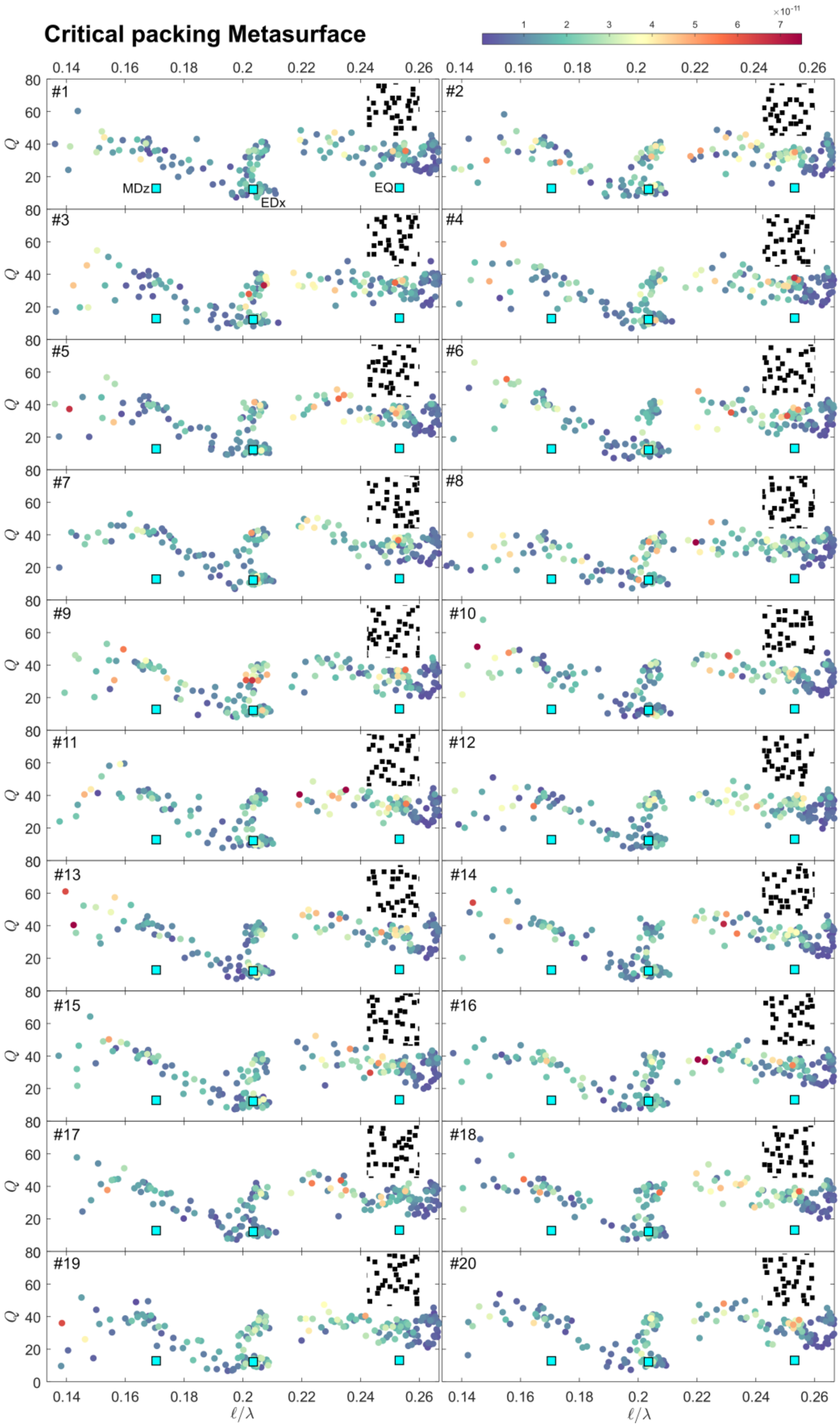

**FIG. S2.3** Same as in Fig. S2 for metasurfaces operating at critical packing. Parameters are the same as in Fig. 3a(ii).

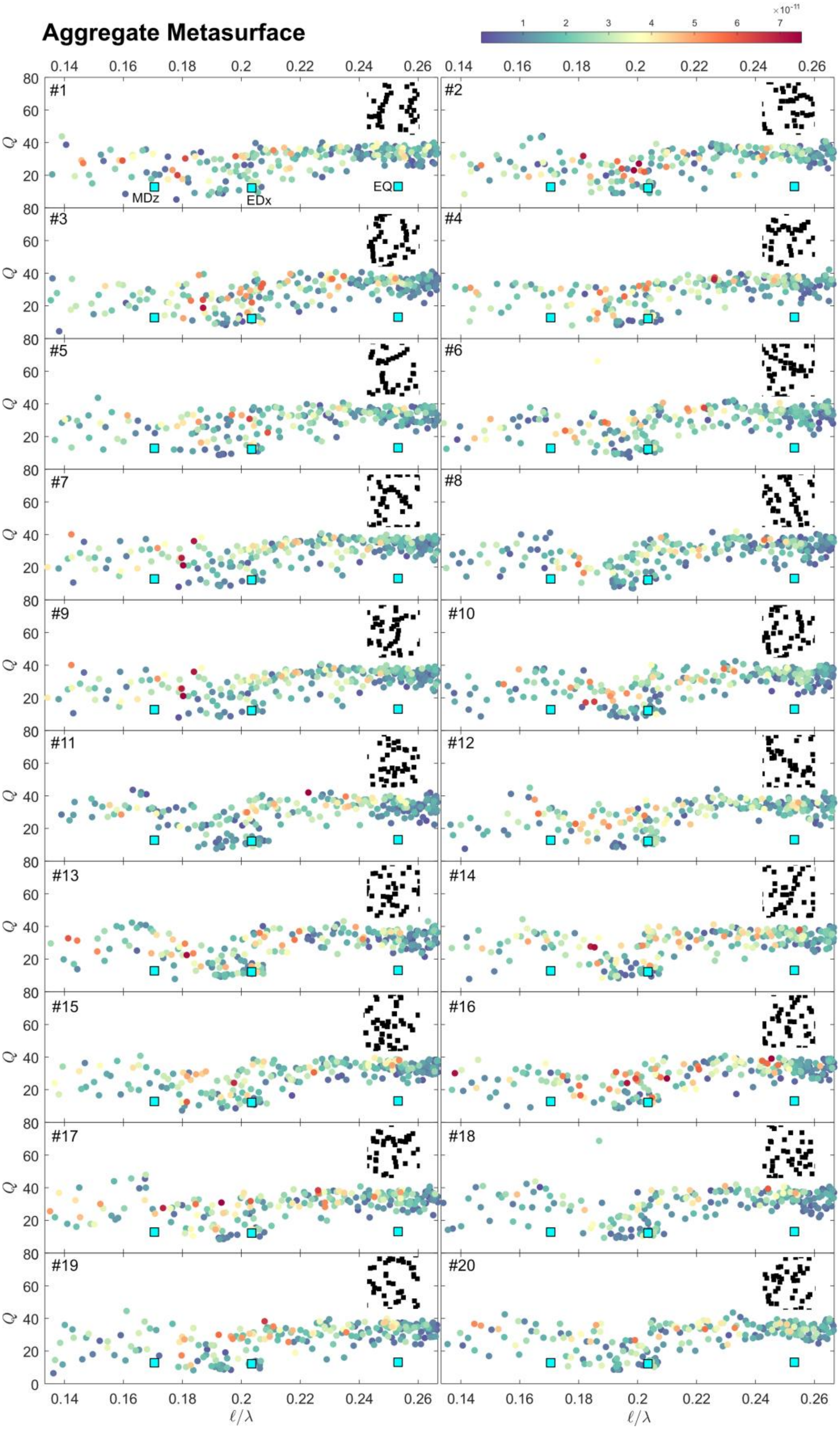

**FIG. S2.4** Same as in Fig. S2 for aggregate metasurfaces. Parameters are the same as in Fig. 3a(iv).

## S2.3 Odd symmetry QNMs at critical packing

The statistical variations observed for odd symmetry QNMs are similar to those observed in Section S2.2 for even symmetry QNMs. We just display the critical packing case hereafter in Fig. S2.5 for 20 independent disorder realizations. The corresponding DOS map obtained with 40 independent realizations is shown in Fig. S2.6.

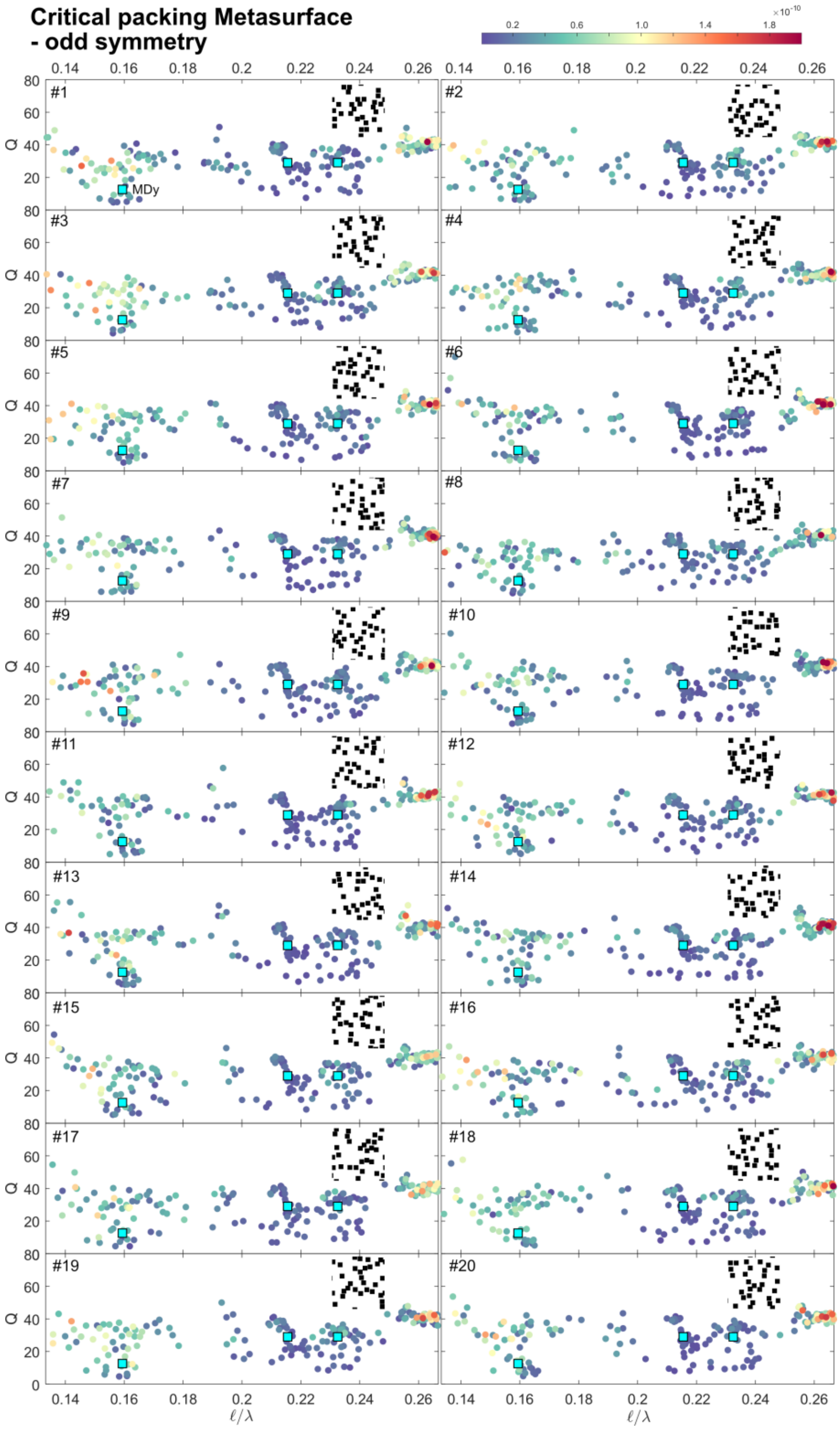

**FIG. S2.5** Complex eigenfrequency maps of odd-symmetric QNMs for 20 independent disorder realizations of metasurfaces operating at critical packing. The parameters of all metasurfaces are the same as in Fig. 3a(ii). The insets illustrate the distribution of meta-atoms within the supercell.

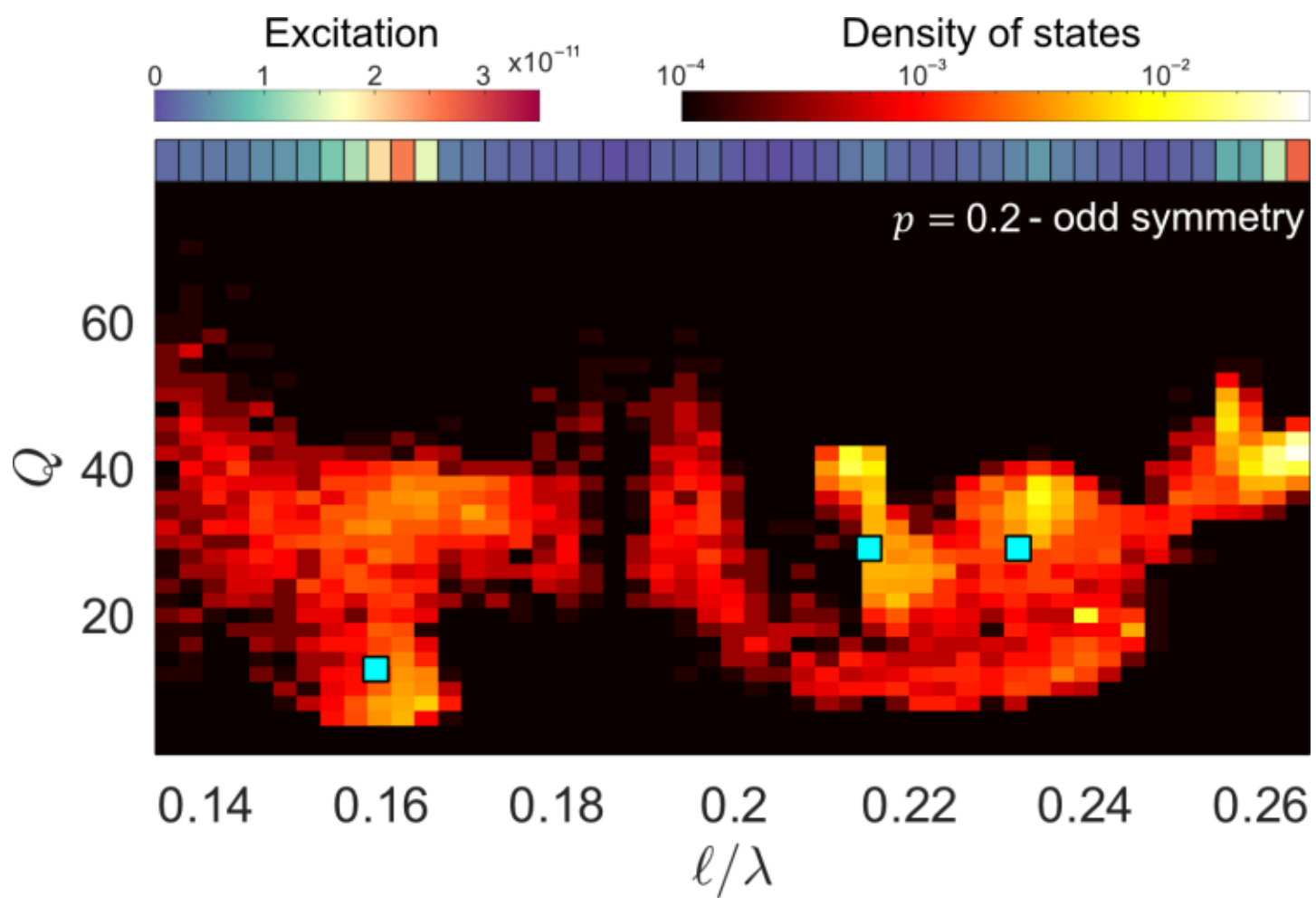

**FIG. S2.6** DOS maps computed from 40 statistically independent realizations, with their complex-frequency planes shown in Fig. S2.5, for odd symmetry QNMs at critical packing. The three bright blue squares indicate the dominant modes of a single nanopost. The colour bands above each map indicate QNM excitation probability densities. A close resemblance with Fig. 3(ii) for the even modes is noticeable.

### S2.4 Convergence test for computations performed for larger supercells

In this section, we compare the QNM maps previously computed using 1.2×1.2 μm$^2$ supercells with new QNM maps computed for four times larger (2.4×2.4 μm$^2$) supercells. While the geometries of the two supercells differ significantly, they both follow the same statistical distribution of metaatom disorder. Specifically, the silicon filling fraction is fixed at 0.2 (critical packing), and the packing fraction remains $p = 0.2$, like in Fig. 3a(ii).

We focus on even-symmetric QNMs. Figure S2.7(i)–(iv) display the complex-frequency maps for four independent realizations computed for the smaller supercells. These realizations, extracted from Fig. S2.3, are combined in Fig. S2.7(v). Each complex-frequency map for the small supercells contains approximately 190 QNMs, while the corresponding map for the larger supercell in Fig. S2.7(vi) contains around 750 QNMs.

There is no direct correspondence between individual realizations of small and large supercells beyond statistical correlations. However, the distribution of QNMs in the complex plane, along with the distribution and magnitude of excitation coefficients, remains qualitatively similar. This consistency reinforces confidence that the statistical analysis conducted with the 1.2×1.2 μm$^2$ supercells in Fig. 3a is sufficiently representative for comparing the DOS across particulate, critical-packing, and aggregate metasurfaces.

To enable comparison between Figs. S2.7(v) and S2.7(vi), the excitation coefficients $|\alpha_m| \sim |2(\varepsilon_{Si} - \varepsilon_b) Q_m \iiint \mathbf{E}_b \cdot \tilde{\mathbf{E}}_m \, d^3r$ (see Methods) in Fig. S2.7(vi) have been divided by 2. This adjustment accounts for two compensating factors arising from the increase in supercell size. When the unit cell expands from 1.2 μm to 2.4 μm, the number of nanoposts increases by a factor of four. Simultaneously, the normalized QNM electric field in the integrand is reduced by half due to QNM normalization [4], leading to a net factor of 2 in the integral.

The computation of 1000 Eigenvectors with MAN takes around 120 minutes on a PC computer equipped with 2.40 GHz ×2 processors for 1.2×1.2 μm$^2$ supercell size. The computation of 3000 Eigenvectors for 2.4×2.4 μm$^2$ supercell size lasts around 1,400 minutes. The times of the QNM computations are mainly determined by the number of degrees of freedom that is proportional to the number of mesh elements and the number of eigenmodes to be computed. The number of degrees of freedom is around $3 \times 10^5$ for critical packing samples with 1.2×1.2 μm$^2$ supercell size, and around $1.5 \times 10^6$ for critical packing samples with 2.4×2.4 μm$^2$ supercell size.

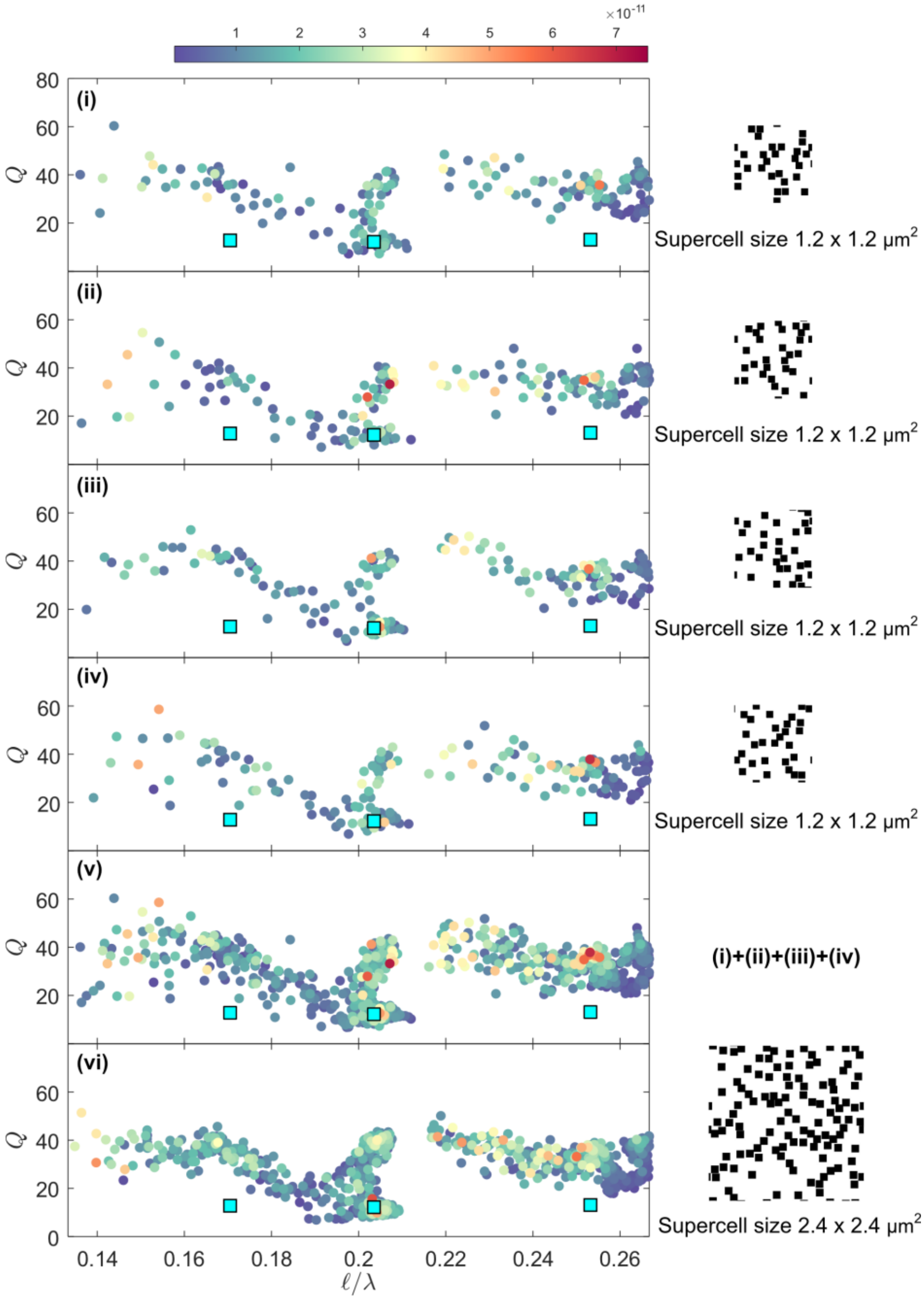

**FIG. S2.7** Comparison between the QNM population and excitation coefficient for metasurfaces of different supercell sizes at critical packing. (**i**)-(**iv**) display four complex-frequency planes computed for 4 independent realisations obtained with 1.2×1.2 μm$^2$ supercells. (**v**) gathers all the QNMs of (**i**)-(**iv**) in one plot. (**vi**) display a complex-frequency plane computed for a single realisation obtained with a 2.4×2.4 μm$^2$ supercell. The colorbar shows the QNM excitation coefficients for a normally incident plane wave.

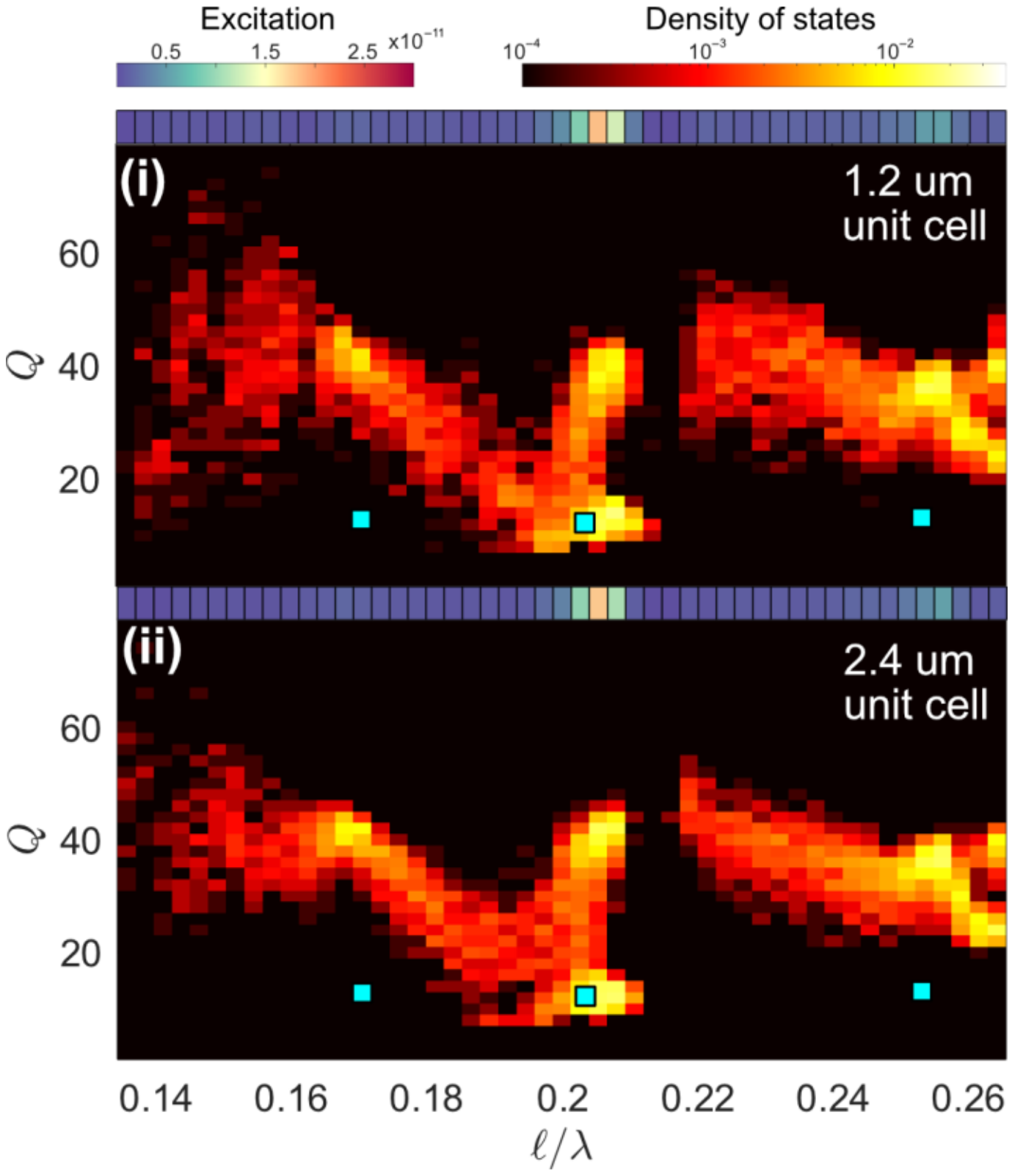

**FIG. S2.8** Comparison of the DOS maps of even electromagnetic modes of critical packing metasurface computed from **(i)** 1.2×1.2 μm$^2$ supercell and **(ii)** 2.4×2.4 μm$^2$ supercell. The colour bands above each map indicate QNM excitation probability densities. Note that (**i**) is averaged over 40 independent realizations and (**ii**) is averaged over 10 independent realizations.

Figure S2.8 presents the comparison of the DOS maps of even modes at critical packing computed using supercells of 1.2×1.2 μm$^2$ and 2.4×2.4 μm$^2$. Figure S2.8(i) is obtained by averaging over 40 independent realizations, while Fig. S2.8(ii) is obtained using 10 independent realizations. This difference arises because the 1.2×1.2 μm$^2$ stypically contain approximately four times more QNMs than the 2.4×2.4 μm$^2$ supercells. Despite this variation, the QNM DOS and excitation probability densities, depicted as colour bands, remain qualitatively similar across both computations. This consistency further validates the statistical analysis performed with 1.2×1.2 μm$^2$ supercells in Fig. 3a.

### S2.5 Complex frequency maps of the selected near-field distributions in Fig. 3b

For accuracy, we provide the complex-frequency maps for the QNM near-field distributions shown in Fig. 3b. These maps belong to Figs. S2.2-4.

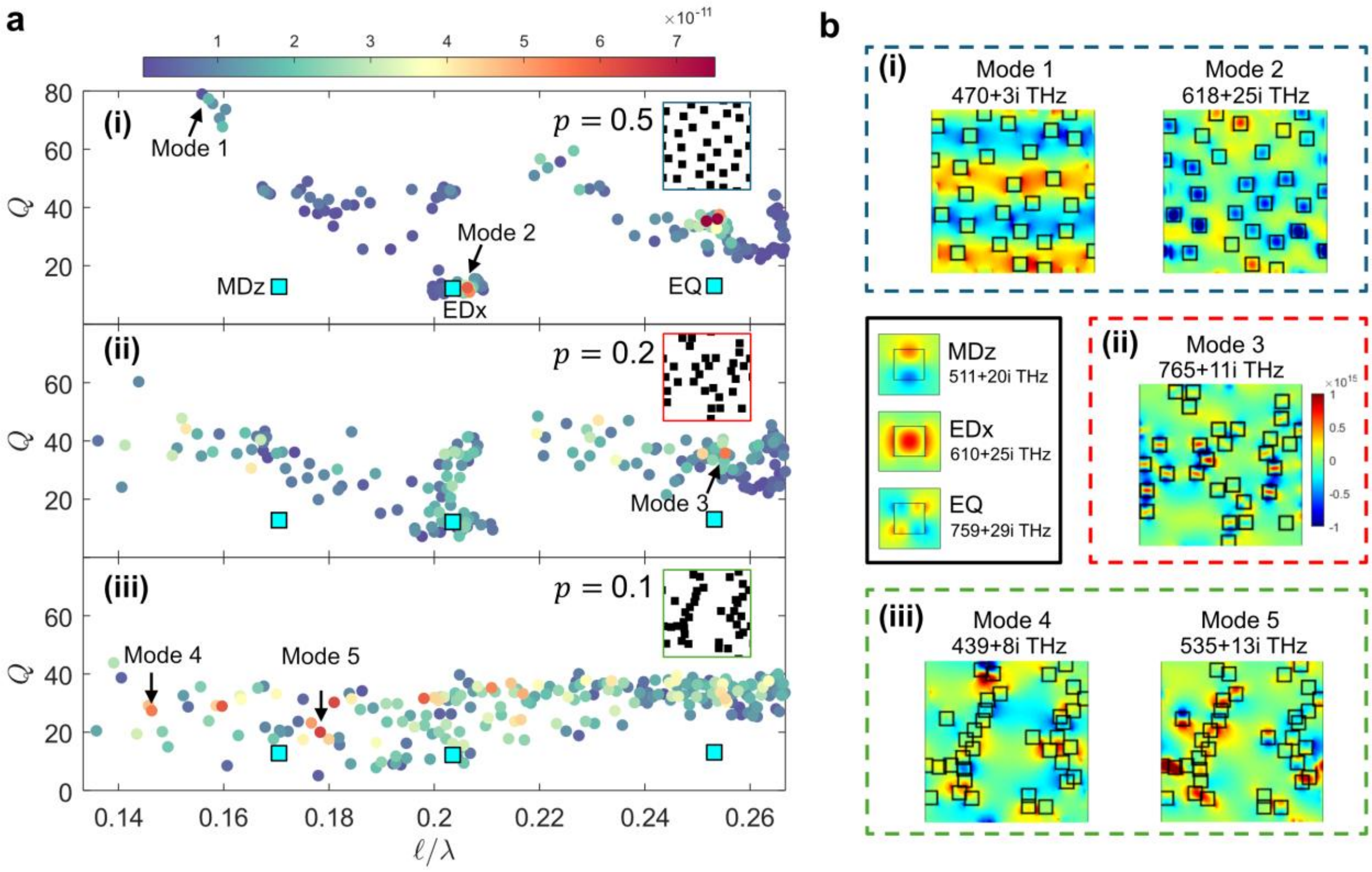

**FIG. S2.9 a** Complex frequency maps for the QNM near-field distributions shown in Fig. 3b. The colour of the QNM dots represents the absolute value of their excitation coefficient. Insets depict the metasurfaces in the $xy$-plane, where black represents silicon and white denotes air. **b** Same as Fig. 3b in the main text. Note that modes 1-5 are indicated with black arrows in **a**.

## S3. Diffuse BRDF model to predict the blue-shift effect

The diffuse BRDF model in the main text (Eq. (3)) separates the contributions of individual nanoparticles and structural correlations to the diffuse intensity. It combines full-wave electromagnetic simulations of the scattering by an individual Si nanopost on a silica substrate with perturbation theory [5] to account for multiple scattering with a correction term [2].

In the present work, the model is adapted to incorporate experimental measurements of the specular reflectance $R_s$ and transmittance $T_s$. Hereafter, we provide the flow chart for computing the correction factor $C(\mathbf{k}_s, \hat{\mathbf{e}}_s, \mathbf{k}_i, \hat{\mathbf{e}}_i)$ and then the diffuse BRDF $f_{\text{diff}}(\mathbf{k}_s, \hat{\mathbf{e}}_s, \mathbf{k}_i, \hat{\mathbf{e}}_i)$ used to compute the BRDFs in Fig. 4.

The first step is to measure $T_s(\theta_i, \hat{\mathbf{e}}_i)$ and $R_s(\theta_i, \hat{\mathbf{e}}_i)$ for all incidence angles $\theta_i$ and polarizations $\hat{\mathbf{e}}_i$ as a function of incident wave frequency (Fig. S5a).

Next, a full-wave simulation using COMSOL is performed to compute the extinction cross-section $\sigma_e(\theta_i, \hat{\mathbf{e}}_i)$ of an individual Si nanopost on the silica substrate (Fig. S5b).

Once $T_s$, $R_s$ and $\sigma_e$ are determined, the coefficient $\gamma(\theta_i, \hat{\mathbf{e}}_i)$ is calculated as

$$1 - T_s(\theta_i, \hat{\mathbf{e}}_i) - R_s(\theta_i, \hat{\mathbf{e}}_i) \approx \gamma(\theta_i, \hat{\mathbf{e}}_i) \frac{\rho \sigma_e(\theta_i, \hat{\mathbf{e}}_i)}{\cos(\theta_i)}. \tag{S1}$$

The correction factor is then determined as the harmonic mean of the excitation coefficients for the incident and scattered plane waves (Fig. S5c) [2]

$$C(\mathbf{k}_s, \hat{\mathbf{e}}_s, \mathbf{k}_i, \hat{\mathbf{e}}_i)^{-1} \approx \frac{1}{2}\left(\gamma(\theta_s, \hat{\mathbf{e}}_s)^{-1} + \gamma(\theta_i, \hat{\mathbf{e}}_i)^{-1}\right). \tag{S2}$$

Under ISA, the diffuse component of the BRDF for TE and TM polarizations is given by (Fig. S5f)

$$f_{\rm diff}({\rm ISA}) \approx \rho \frac{d\sigma_{\rm s}^{(1)}}{d\Omega}(\mathbf{k}_{\rm s}, \hat{\mathbf{e}}_{\rm s}, \mathbf{k}_{\rm i}, \hat{\mathbf{e}}_{\rm i}) S_{\rm r}\left(\mathbf{k}_{{\rm s},\parallel}, \mathbf{k}_{{\rm i},\parallel}\right) \frac{1}{\cos(\theta_{\rm i})\cos(\theta_{\rm s})}. \quad \text{(S3)}$$

The single-particle scattering diagram $d\sigma_{\rm s}^{(1)}/d\Omega$ is numerically computed with a near-to-far-field freeware [6] coupled to COMSOL Multiphysics (Fig. S5d). The structure factor $S_{\rm r}$, which describes spatial correlations between pairs of particles, is computed with the semi-analytical Baus–Colot model for hard-disk liquids [7] (Fig. S5e).

The final diffuse BRDF is computed by multiplying $f_{\rm diff}({\rm ISA})$ and $C(\mathbf{k}_{\rm s}, \hat{\mathbf{e}}_{\rm s}, \mathbf{k}_{\rm i}, \hat{\mathbf{e}}_{\rm i})$ and averaging over TE and TM polarizations (Fig. S5g)

$$f_{\rm diff} \approx \rho \frac{d\sigma_{\rm s}^{(1)}}{d\Omega}(\mathbf{k}_{\rm s}, \hat{\mathbf{e}}_{\rm s}, \mathbf{k}_{\rm i}, \hat{\mathbf{e}}_{\rm i}) S_{\rm r}\left(\mathbf{k}_{{\rm s},\parallel}, \mathbf{k}_{{\rm i},\parallel}\right) \frac{C(\mathbf{k}_{\rm s}, \hat{\mathbf{e}}_{\rm s}, \mathbf{k}_{\rm i}, \hat{\mathbf{e}}_{\rm i})}{\cos(\theta_{\rm i})\cos(\theta_{\rm s})}. \quad \text{(S4)}$$

A comparison between Figs. S5g and S5h reveals that the blue shift of the diffuse light is not captured when multiple scattering effects are neglected. Furthermore, when compared with the measured BRDF in Fig. 4(c), the ISA model tends to overestimate the diffuse light intensity—a common limitation of ISA models [8]. The correction factor primarily increases the diffuse intensity for wavelengths $\lambda < 450$ nm while reducing it around $\lambda = 500$ nm, particularly at high angles.

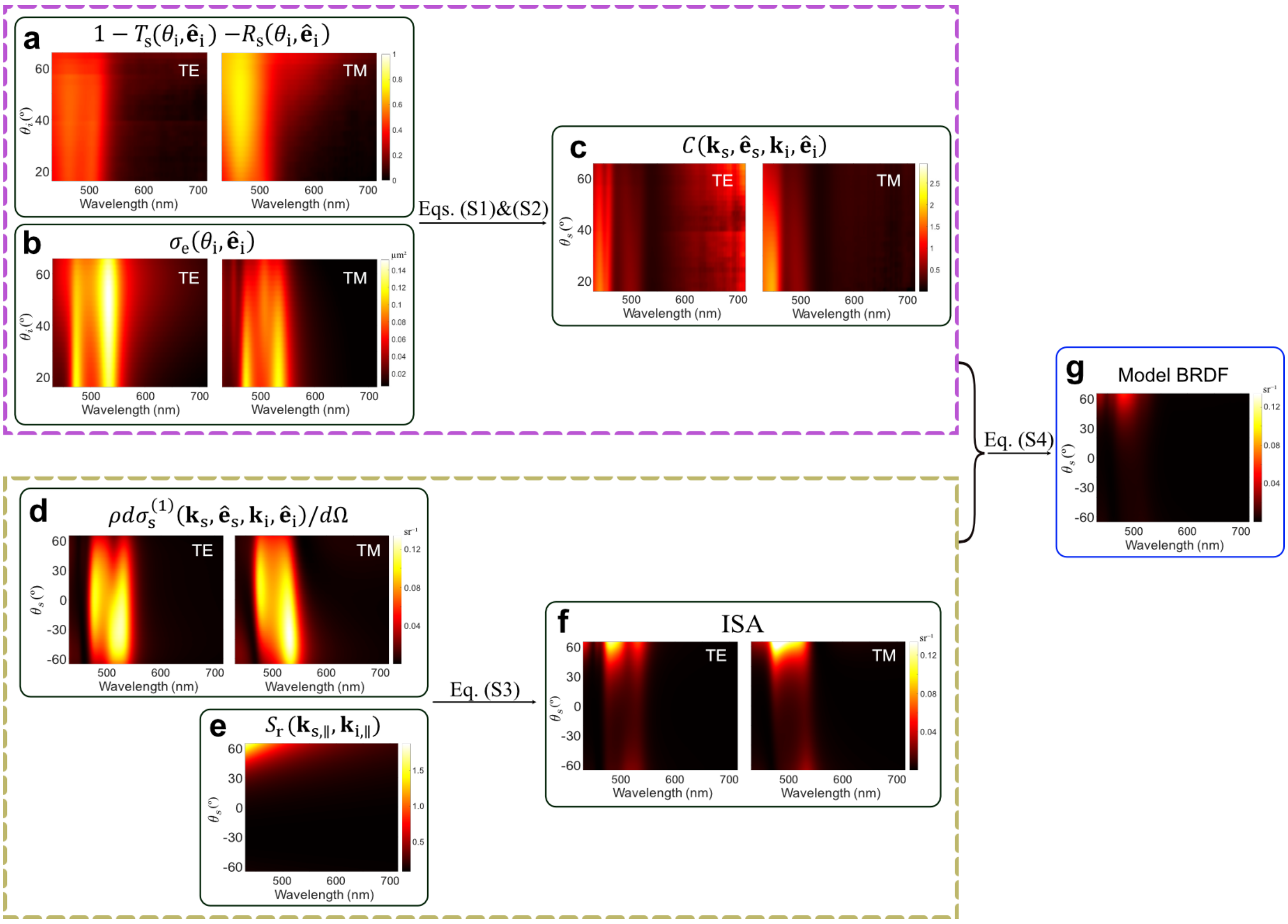

**FIG. S3** Flow Chart of applying the diffuse BRDF model in the main text.

## S4. Extended Maxwell-Garnet model

The electromagnetic homogenization of disordered assemblies of particles (also known as particulate composites) has been a long-standing topic that started more than a century ago and has experienced many developments over the years. The properties of the effective medium are traditionally obtained with the help of so-called mixing laws which relate the relevant macroscopic parameters of the ensemble to the knowledge of the properties of the constituent materials making up the inclusions and the host material.

In this section, we compare our measurement on specular reflection of particulate and critical packing samples to the Extended Maxwell Garnett mixing formula

$$\varepsilon_{EMG} = \frac{x^3+3ifa_1}{x^3-\frac{3}{2}ifa_1}, \mu_{EMG} = \frac{x^3+3ifb_1}{x^3-\frac{3}{2}ifb_1}, \quad \text{(S5)}$$

where $x = \sqrt{\varepsilon_b}\omega a/c$, $a$ is the radius of sphere, $f$ is the filling fraction of scatterers, $a_1$ and $b_1$ are Mie coefficient of single metaatom.

The effective permittivity and permeability computed with Eq. (S5) are then incorporated into a 4×4 transfer matrix method using the freeware RETICOLOfilm-stack [9] to compute the specular reflection spectrum by taking account the glass substrate.

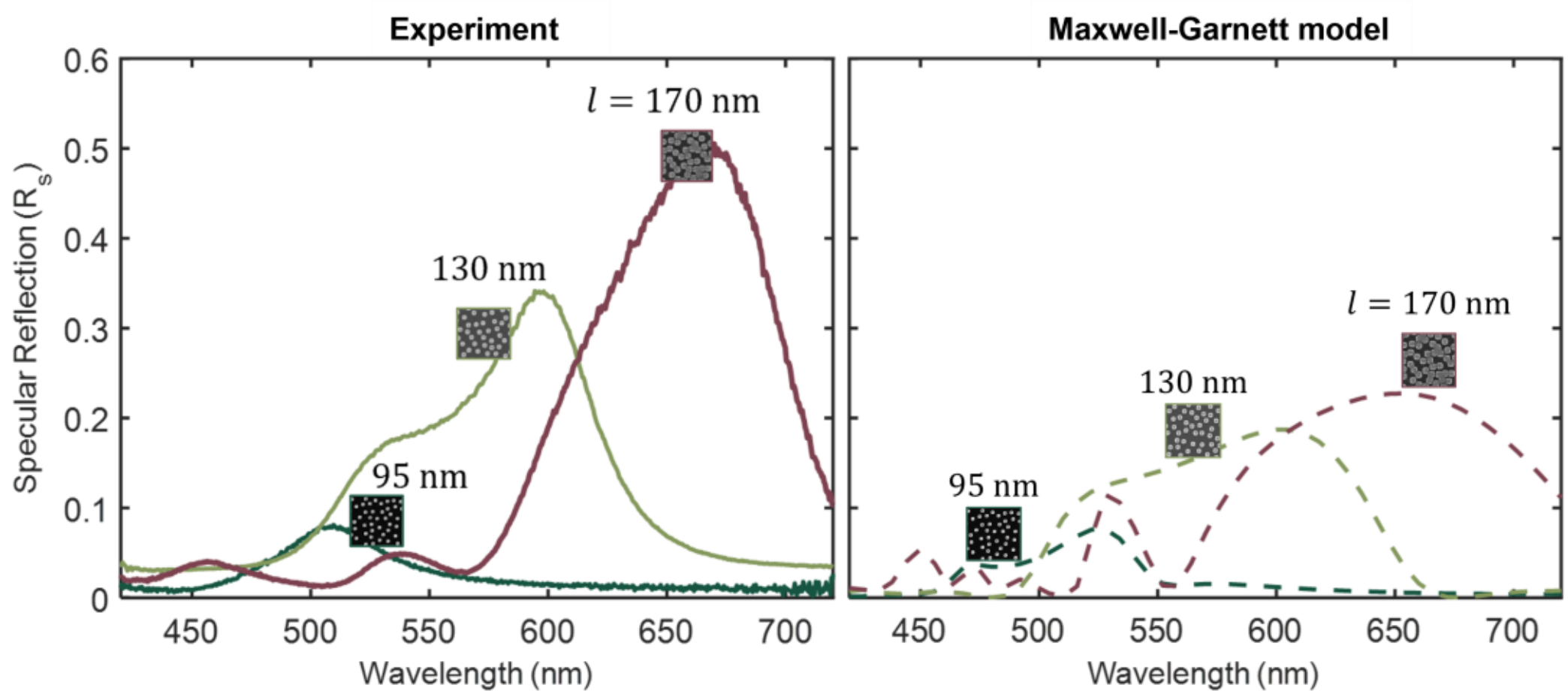

**FIG. S4** Specular reflectance spectra of two particulate metasurfaces and one critical-packing metasurface. Left: experiment. Right: Maxwell-Garnett model.

The specular reflection of particulate metasurfaces is well approximated by the Maxwell-Garnett mixing formula. However, for critical packing sample, the model becomes inaccurate as expected. It predicts a much lower reflection peak compared to measurement.

## S5. Reliability of the ping-pong ball setup

The ping-pong ball serves as a hemi-spherical projection screen to visualize the diffuse light. Goodling, A. E. et. al., has used half of ping-pong ball to image the reflecting iridescent colour of oil droplets [10]. Kim, M., has used it to image the reflected diffraction light of a periodic plasmonic structure [11]. In these works, the ping-pong ball is simply considered to be translucent.

Fig.S5.1 presents the measured transmission spectra of the half ping-pong ball at different observation angles. It demonstrates a slight preference for scattering red light more efficiently than blue, while maintaining a relatively flat scattering efficiency across the visible spectrum. This uniform scattering ensures that light is diffused evenly, minimizing hotspots or colour imbalance. The near-flat scattering response means that the film performs consistently across a range of wavelengths, making it ideal for use with white light or full-spectrum projection. Its ability to disperse light efficiently in all directions enhances image clarity and brightness, providing a smooth surface for high-quality visual displays.

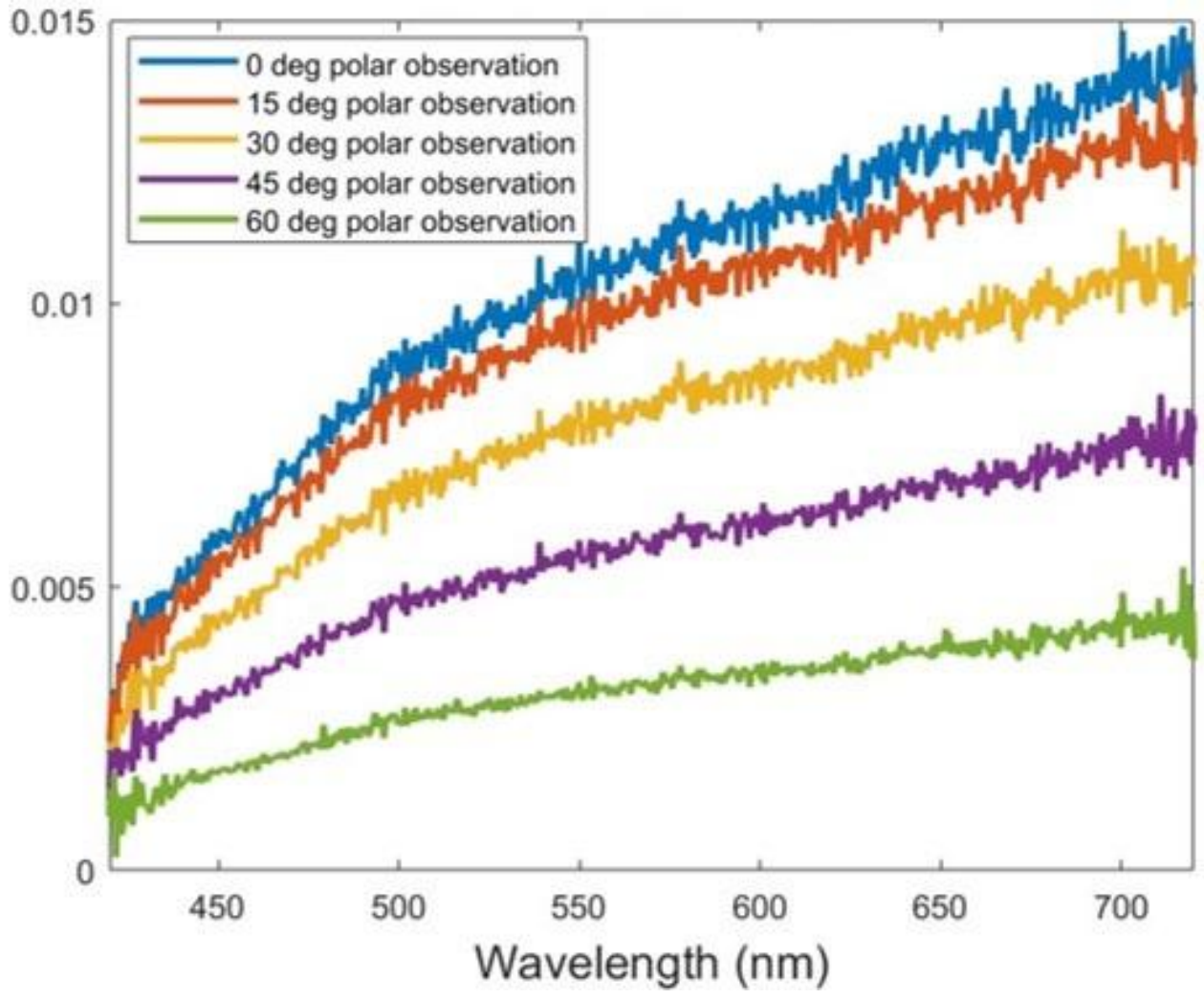

**FIG. S5.1** Transmission spectra of the half ping-pong ball at different observation angles

The measured spectrum of the illuminating supercontinuum laser is shown in Fig. S5.2. This spectrum is different from the Standard illuminant D65. However, its spectrum is similar to a typical white LED light [12].

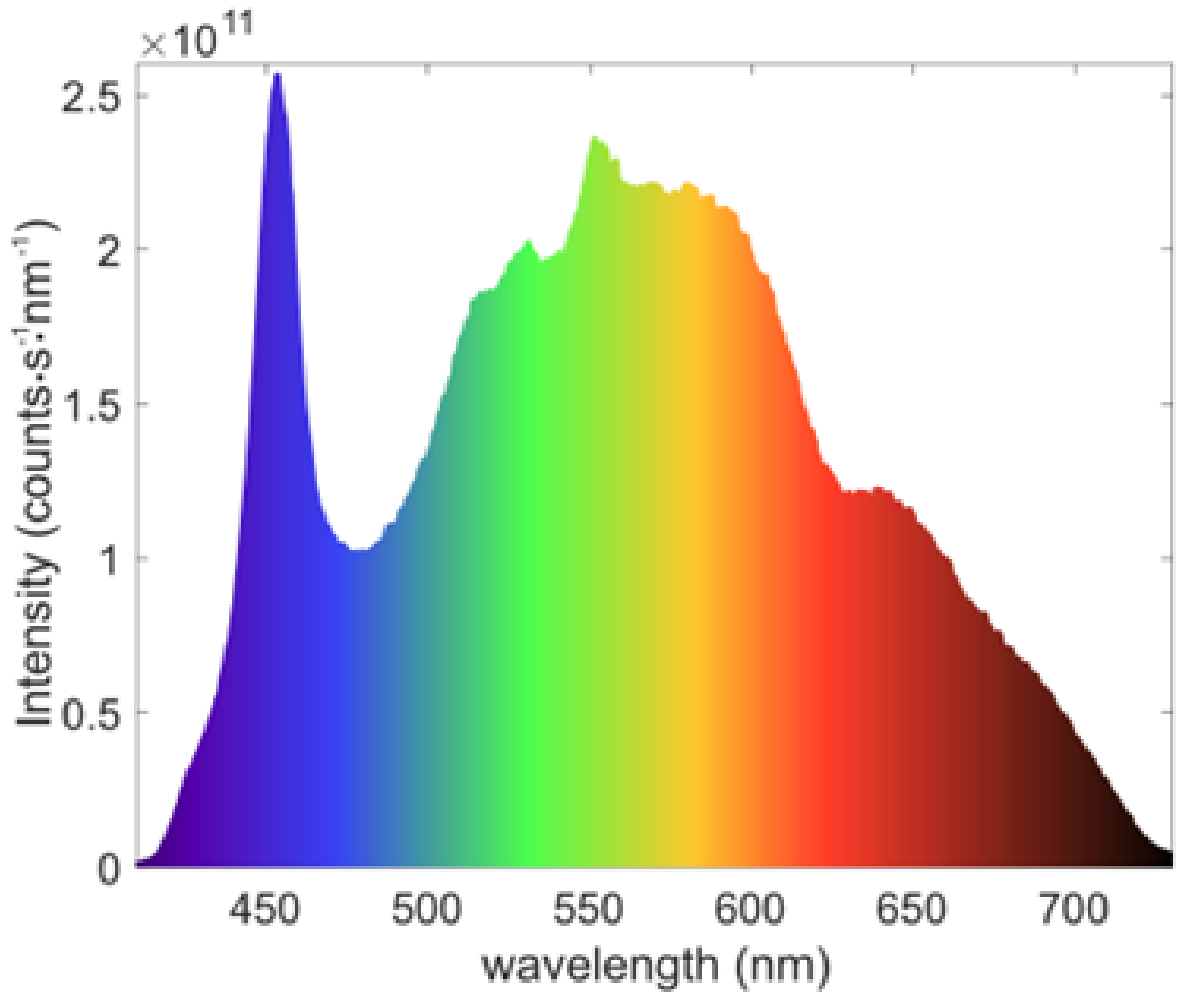

**FIG. S5.2** Input Supercontinuum laser spectrum.

The experimental setups for taking the ping-pong ball (diffuse light) pictures and specular light pictures are shown in FIG. S5.3. The supercontinuum laser passes through a beam expander, an iris and a lens to focus onto the metasurface. The diffuse light and specular light pictures are recorded with Canon EOS 1000D camera mounted on an optical table at fixed exposure times for each case. For taking the specular light pictures, a diffuser was positioned between the metasurface sample and the camera.

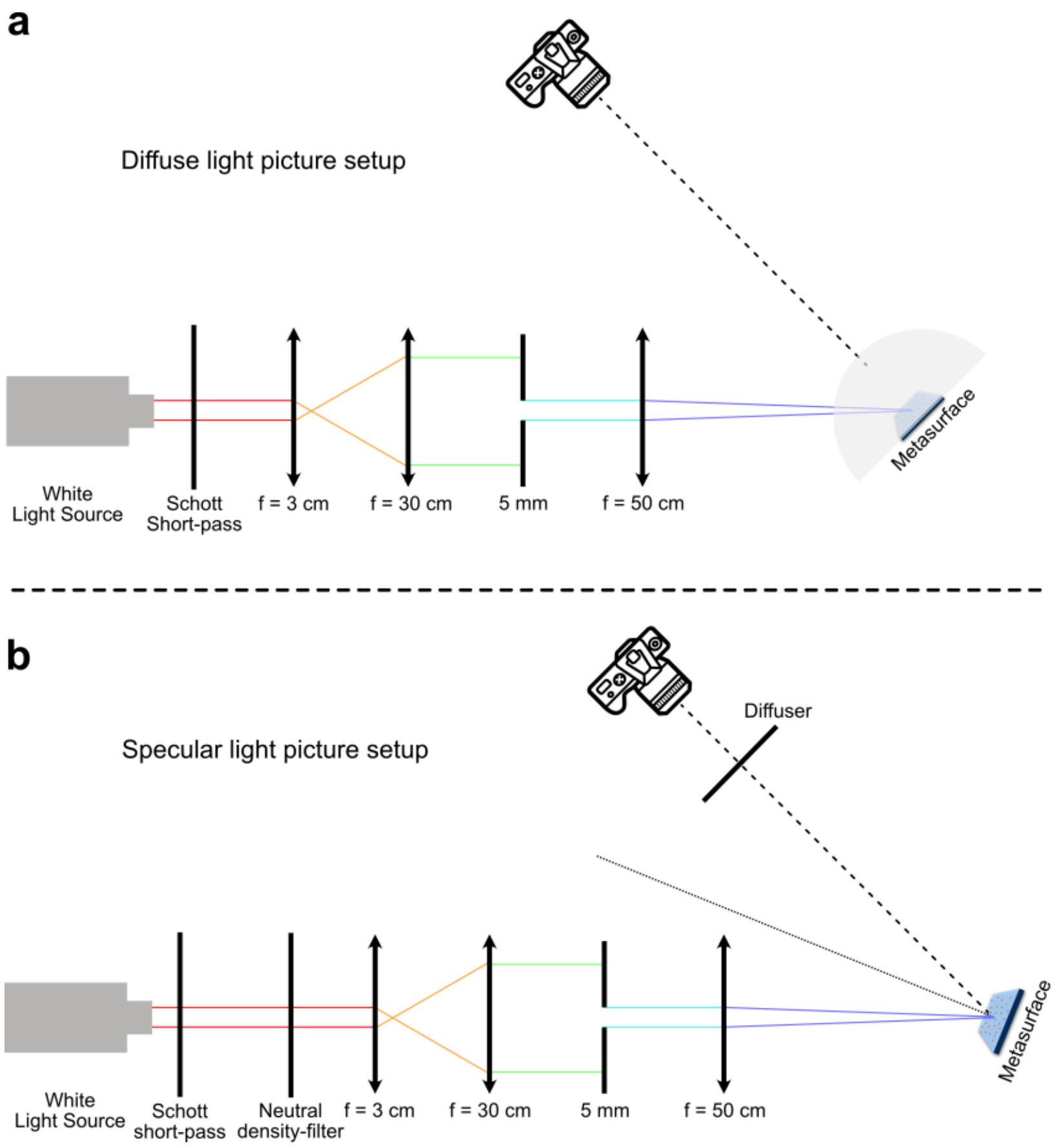

**FIG. S5.3** Experimental setup for taking pictures of **a** diffuse light and **b** specular light.

## S6. Diffuse colour in CIELAB map

To complement the CIE 1931 chromaticity diagram presented in Fig. 5b, we show in Supplementary Fig. S6 the same five diffuse colours plotted in the CIE 1976 L*a*b* colour space. In this representation, the vertical axis (L*) directly quantifies brightness, while the horizontal axes (a* and b*) describe the green–red and blue–yellow components of the colour, respectively. The conversion from the measured CIE 1931 XYZ coordinates to CIE 1976 L*a*b* was performed using the D65 reference white point. This 3D view provides a more comprehensive representation of the colour appearance, highlighting both the intensity and the hue of the diffuse scattering.

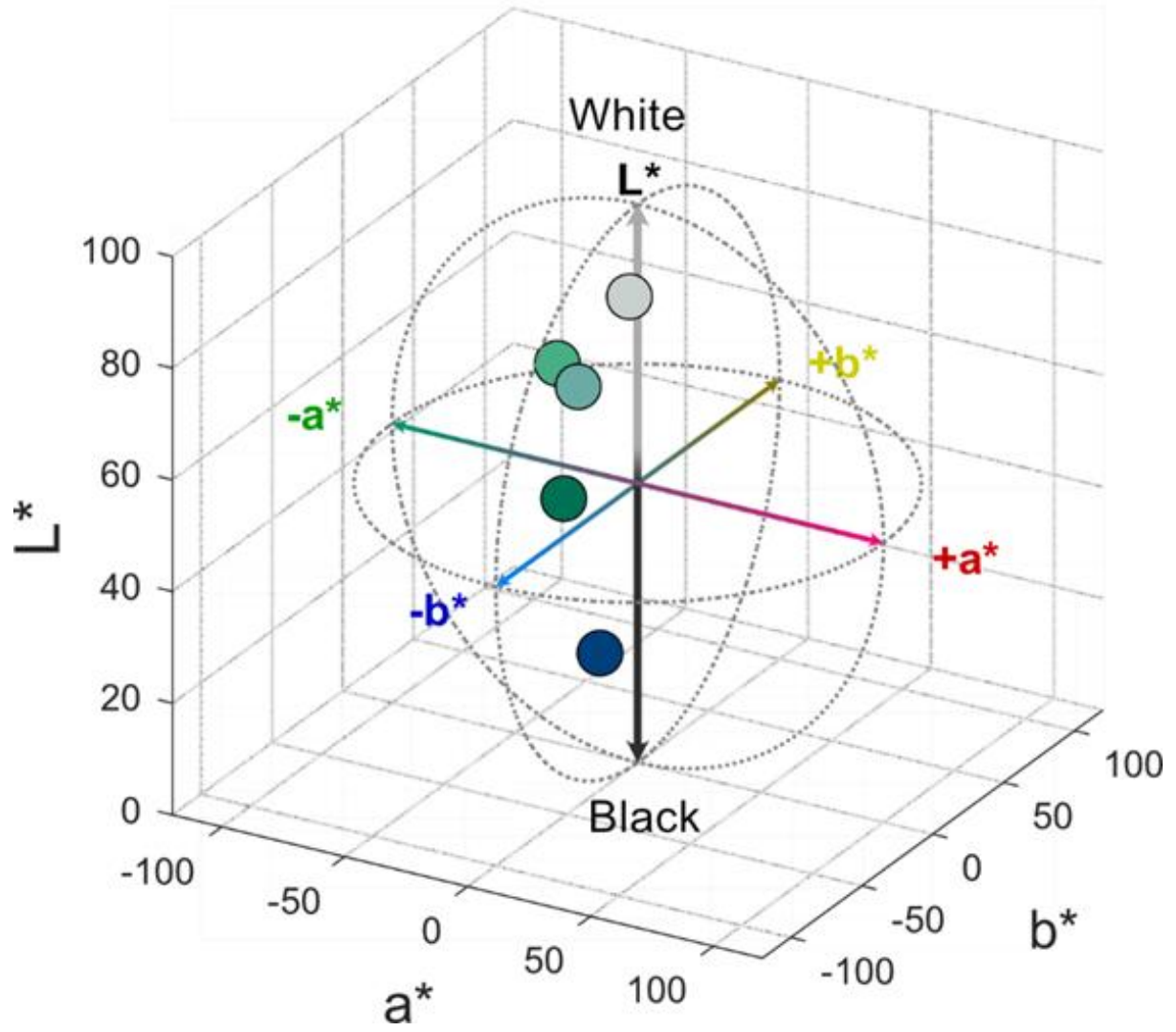

**FIG. S6** Diffuse colours from Fig. 5 represented in the CIE 1976 L*a*b* colour space. This 3D representation explicitly includes brightness (L*) in addition to chromaticity (a*, b*), thereby complementing the 2D chromaticity plot shown in Fig. 5b. The conversion from CIE 1931 XYZ coordinates was performed using the D65 reference white point.

## References

1. P. M. Piechulla, B. Fuhrmann, E. Slivina, et al., "Tailored light scattering through hyperuniform disorder in self-organized arrays of high-index nanodisks", Adv. Optical Mater. **9**, 2100186 (2021).
2. K. Vynck, R. Pacanowski, A. Agreda, A. Dufay, X. Granier, P. Lalanne, "The visual appearances of disordered optical metasurfaces", Nat. Mater. **21**, 1035-41 (2022).
3. T. Wu, D. Arrivault, W. Yan, P. Lalanne, "Modal analysis of electromagnetic resonators: user guide for the MAN program", Comput. Phys. Commun. **284**, 108627 (2023).
4. C. Sauvan, J.-P. Hugonin, I. S. Maksymov and P. Lalanne, "Theory of the spontaneous optical emission of nanosize photonic and plasmon resonators", Phys. Rev. Lett **110**, 237401 (2013).
5. L. Tsang and J. A. Kong, *Scattering of electromagnetic waves: advanced topics*, (John Wiley & Sons, 2004).
6. J. Yang, J.-P. Hugonin, and P. Lalanne, "Near-to-far field transformations for radiative and guided waves", ACS Photon. **3**, 395–402 (2016).
7. M. Baus and J.-L Colot, "Thermodynamics and structure of a fluid of hard rods, disks, spheres, or hyperspheres from rescaled virial expansions", Phys. Rev. A **36**, 3912 (1987).
8. B. X. Wang and C. Y. Zhao, "The dependent scattering effect on radiative properties of micro/nanoscale discrete disordered media", Annu. Rev. Heat Transfer **23**, 231-353 (2020).
9. J.-P. Hugonin and P. Lalanne, RETICOLOfilm-stack, DOI: 10.5281/zenodo.7429030
10. A. E. Goodling, S. Nagelberg, B. Kaehr, et al., "Colouration by total internal reflection and interference at microscale concave interfaces", Nature **566**, 523–527 (2019)
11. M. Kim, J. B. Kim, S. Y. Lee, et al., "Designing Multicolor Graphics of Plasmonic Metasurfaces through Gradual Protrusion of Particles at Free Interface", Adv. Mater. Interfaces **9**, 2102240 (2022).
12. https://zeiss-campus.magnet.fsu.edu/print/lightsources/leds-print.html